\documentclass[a4paper,fleqn]{cas-dc}

\usepackage{xcolor} 
\newcommand{\black}[1]{\textcolor{black}{#1}} 
\usepackage[authoryear]{natbib}
\usepackage{array}
\usepackage{booktabs}
\usepackage{caption}
\usepackage{adjustbox}
\usepackage{makecell} 
\usepackage{tabularx} 
\usepackage{ragged2e} 
\usepackage{multirow}
\usepackage{threeparttable}
\usepackage{color}

\def\tsc#1{\csdef{#1}{\textsc{\lowercase{#1}}\xspace}}
\tsc{WGM}
\tsc{QE}

\begin{document}
\let\WriteBookmarks\relax
\def\floatpagepagefraction{1}
\def\textpagefraction{.001}

\shorttitle{}  

\shortauthors{Jinquan Guan, et al.}  

\title[mode = title]{A High Magnifications Histopathology Image Dataset for \\ Oral Squamous Cell Carcinoma Diagnosis and Prognosis}  




\author[1]{\black{Jinquan Guan}}\cormark[1]
\author[2]{\black{Junhong Guo}}\cormark[1]
\author[4]{\black{Qi Chen}}
\author[1]{\black{Jian Chen}} 
\author[2]{\black{Yongkang Cai}}
\author[2]{\black{Yilin He}}
\author[2]{\black{Zhiquan Huang}}  
\author[2]{\black{Yan Wang}}\cormark[2]  
\author[3]{\black{Yutong Xie}}\cormark[2]

\cortext[1]{Equal contribution}
\cortext[2]{Corresponding author}

\affiliation[1]{organization={School of Software Engineering, South China University of Technology},
            city={Guangzhou},
            country={China}}

\affiliation[2]{organization={Department of Oral and Maxillofacial Surgery, Sun Yat-sen Memorial Hospital, Sun Yat-sen University},
            city={Guangzhou},
            country={China}}

\affiliation[3]{organization={Mohamed Bin Zayed University of Artificial Intelligence},
            city={Abu Dhabi},
            country={UAE}}

\affiliation[4]{organization={School of Computer Science, University of Adelaide},
            city={Adelaide},
            country={Australia}}
            



\begin{abstract}
Oral Squamous Cell Carcinoma (OSCC) is a prevalent and aggressive malignancy where deep learning-based computer-aided diagnosis and prognosis can enhance clinical assessments.
However, existing publicly available OSCC datasets often suffer from limited patient cohorts and a restricted focus on either diagnostic or prognostic tasks, limiting the development of comprehensive and generalizable models.
To bridge this gap, we introduce Multi-OSCC, a new histopathology image dataset comprising 1,325 OSCC patients, integrating both diagnostic and prognostic information to expand existing public resources.
Each patient is represented by six high resolution histopathology images captured at $\times$200, $\times$400, and $\times$1000 magnifications—two per magnification—covering both the core and edge tumor regions.
The Multi-OSCC dataset is richly annotated for six critical clinical tasks: recurrence prediction (REC), lymph node metastasis (LNM), tumor differentiation (TD), tumor invasion (TI), cancer embolus (CE), and perineural invasion (PI). To benchmark this dataset, we systematically evaluate the impact of different visual encoders, multi-image fusion techniques, stain normalization, and multi-task learning frameworks.
Our analysis yields several key insights: 
(1) The top-performing models achieve excellent results, with an Area Under the Curve (AUC) of 94.72\% for REC and 81.23\% for TD, while all tasks surpass 70\% AUC. 
(2) Stain normalization benefits diagnostic tasks but negatively affects recurrence prediction; 
(3) Multi-task learning incurs a 3.34\% average AUC degradation compared to single-task models in our multi-task benchmark, underscoring the challenge of balancing multiple tasks in our dataset.
To accelerate future research, we publicly release the Multi-OSCC dataset and baseline models at \href{https://github.com/guanjinquan/OSCC-PathologyImageDataset}{github.com/guanjinquan/OSCC-PathologyImageDataset}.
\end{abstract}


\maketitle

\section{Introduction}
\label{sec:introduction}

Oral Squamous Cell Carcinoma~(OSCC) is a common malignant head and neck tumour. According to global cancer statistics, more than 380,000 patients with oral cancer were diagnosed in 2022, of which approximately 180,000 died \citep{bray2024global}.  Accurate diagnosis, effective treatment, and a well-informed prognosis plan are essential to reduce mortality rates. Histopathology checking is a gold standard for identifying OSCC and its status. To achieve this purpose, it is often necessary for clinicians to carry out a histopathology biopsy of the lesion site of the patient, and the biopsy tissues are processed via staining and microtomy to generate histopathology slides. The pathologist confirms the diagnosis through examination and analysis of the histopathology slides, after which clinicians make a more accurate prognosis assessment.

Artificial intelligence~(AI) systems have demonstrated significant potential for rapid and accurate analysis of pathology images~\citep{mckinney2020international}. In the context of OSCC, AI  automated analysis of histopathology images promises to streamline the diagnostic process, enabling precise and efficient identification and classification of cancerous tissues, and ultimately improving patient prognosis~\citep{warin2024deep}. Existing OSCC datasets are shown in Table~\ref{tab:dataset_comparison}, which include: the TCGA-HNSC database~\citep{zuley2016cancer} compiles clinical information, radiological data, genomic data, and histopathology images from 528 patients, most of whom have oral cancer aiming for prognosis purpose. \citep{rahman2020histopathological} published a dataset of optical microscopy images for diagnosing normal and OSCC images. The ORCHID dataset~\citep{chaudhary2024high} includes microscopy images of OSCC and oral submucous fibrosis~(OSMF), supporting cell classification studies and providing tumor differentiation~(TD) labels for OSCC images. However, existing OSCC datasets often have limited patient cohort sizes and focus on specific aspects of diagnosis or prognosis. These limitations constrain the range of clinical problems that their developed AI systems can address, while also hindering the development of more generalized and robust models.

\begin{table}[tbp]
    \centering
    \caption{Comparison of Oral Cancer Datasets. \cite{rahman2020histopathological} and ORCHID collected multiple samples from individual patients, thereby generating a substantial number of images.}
    \label{tab:dataset_comparison}
    \adjustbox{width=\columnwidth,center}{%
    \begin{tabular}{l cccc} 
        \toprule
        \multirow{2}{*}{\textbf{Name}} & \textbf{Year} & \textbf{Patients} & \textbf{Samples} & \textbf{Cancer} \\
        \cmidrule(lr){2-5}
        & \multicolumn{4}{c}{\textbf{Description / Task}} \\
        \midrule
        
        
        \multirow{2}{*}{\makecell[l]{TCGA-HNSC \\ \cite{zuley2016cancer} \\ \textbf{Prognosis}}} 
        & 2014 & 528 & - & SCC \\
        \cmidrule(lr){2-5}
        & \multicolumn{4}{p{0.75\columnwidth}}{
            A database for Squamous Cell Carcinoma survival analysis.
        } \\
        \midrule
        
        \multirow{2}{*}{\makecell[l]{\cite{rahman2020histopathological} \\ \textbf{Diagnosis}}}
        & 2020 & 230 & 1224 & OSCC \\
        \cmidrule(lr){2-5}
        & \multicolumn{4}{p{0.75\columnwidth}}{%
            2-class classification of normal and OSCC images.
        } \\
        \midrule
        
        \multirow{2}{*}{\makecell[l]{ORCHID \\ \cite{chaudhary2024high} \\ \textbf{Diagnosis}}}
        & 2024 & 150 & 14705 & OSCC+OSMF\\
        \cmidrule(lr){2-5}
        & \multicolumn{4}{p{0.75\columnwidth}}{%
            2-Stage task: (1) 3-class classification of normal, OSCC, and OSMF images. (2) Classification of OSCC samples from Task-1 into three classes based on tumor differentiation.
        } \\
        \midrule
        
        \multirow{2}{*}{\makecell[l]{Multi-OSCC (Ours) \\ \textbf{Diagnosis+Prognosis}}}
        & - & 1325 & 1325 & OSCC \\
        \cmidrule(lr){2-5}
        & \multicolumn{4}{p{0.75\columnwidth}}{%
            6-tasks including patient level prognosis (2-year tumor recurrence prediction) and diagnosis (tumor status assessment).
        } \\
        \bottomrule
    \end{tabular}%
    }

\end{table}

To advance research in histopathological image analysis, we introduce Multi-OSCC, a novel dataset of Oral Squamous Cell Carcinoma (OSCC) images featuring multiple targets. 
Following the data collection methodology of \cite{chaudhary2024high} and \cite{rahman2020histopathological}, we capture these histopathology images using a microscope at various high magnifications.
This dataset encompasses six tasks related to the diagnosis and prognosis of OSCC, incorporating a larger patient cohort, with detailed descriptions provided in Table~\ref{tab:task_descriptions}. The tasks in our dataset are based on three clinically relevant scenarios designed to assist clinicians in diagnostic and prognostic analysis:
\begin{itemize}
    \item [1.] \textbf{REC}:  
    This task aims to assist clinicians in identifying the risk of tumor recurrence. Based on the recurrence risk predicted by our model, the clinician can formulate an appropriate prognosis plan for patients who have undergone surgical resection.
    
    \item [2.] \textbf{LNM}:  
    This task helps clinicians decide whether further surgical procedures, such as cervical lymph node dissection, are necessary. Using histopathological images obtained through incisional biopsy, our model predicts the probability of lymph node metastasis, reducing unnecessary lymphadenectomy while ensuring high-risk areas are not overlooked.
    
    \item [3.] \textbf{TD, TI, CE, PI}:  
    These tasks assist clinicians in assessing the severity of the tumor. Since tumor staging (T stage) involves lesion size, which can be measured manually, we focus on more granular diagnostic classifications. The excised lesions from surgery are sent to pathologists for examination, and our model helps them diagnose tumor status and make comprehensive pathological assessments.
\end{itemize}

\begin{table}[tbp]  
  \centering
  \caption{Abbreviation and descriptions of Six tasks for oral squamous cell carcinoma.}
  \resizebox{\columnwidth}{!}{
    \begin{tabular}{@{}>{\centering\arraybackslash}p{1.2cm} >{\centering\arraybackslash}p{1cm} p{5.5cm}@{}} 
      \toprule
       \textbf{Application} & \textbf{Abbreviation} & \parbox[t]{5.5cm}{\centering \textbf{Description}} \\
      \midrule
      Prognosis & REC   & \parbox[t]{5.5cm}{\textbf{Recurrence~(2-classes)} : \\ Predicting OSCC tumor recurrence.} \\
      \midrule
      \multirow[t]{5}{*}{Diagnosis} & LNM   & \parbox[t]{5.5cm}{\textbf{Lymph Node Metastasis~(2-classes)} : \\ Predicting Head and Neck lymph node metastasis.} \\
                                 \cmidrule(lr{0pt}){2-3}  
                                 & TD    & \parbox[t]{5.5cm}{\textbf{Tumor Differentiation~(3-classes)} : \\ Assessing tumor differentiation in histopathology images. A label of 0 indicates high differentiation, while a label of 2 indicates low differentiation. Tumors with low differentiation are more severe.} \\
                                 \cmidrule(lr{0pt}){2-3}  
                                 & TI    & \parbox[t]{5.5cm}{\textbf{Tumor Invasion~(2-classes)} : \\ Assessing oral tumor invasion of surrounding tissues.} \\
                                 \cmidrule(lr{0pt}){2-3}
                                 & CE    & \parbox[t]{5.5cm}{\textbf{Cancer Embolus~(2-classes)} : \\ Estimating vascular invasion (cancer cells infiltrating blood vessels).} \\
                                 \cmidrule(lr{0pt}){2-3}
                                 & PI    & \parbox[t]{5.5cm}{\textbf{Perineural Invasion~(2-classes)} : \\ Estimating perineural invasion (cancer cells infiltrating nerve tissues).} \\
      \bottomrule
    \end{tabular}%
  }
  \label{tab:task_descriptions}%
\end{table}

Compared to prior datasets limited to a single task, our dataset enables joint modeling of diagnosis and prognosis, aligning with clinical workflows. It features multi-task labels and histopathology images from multiple tissue slices per patient, offering a comprehensive resource for multi-target analysis. To the best of our knowledge, this is the first publicly available histopathology image dataset specifically designed for OSCC research, with multiple diagnostic and prognostic targets.

We conduct extensive experiments to evaluate various aspects of our dataset, including comparing vision backbones trained with ImageNet versus histopathology-specific pre-trained weights, examining multi-slice feature fusion strategies, assessing the impact of stain normalization, and exploring multi-task learning in histopathology analysis. The findings of our analysis reveal: 
(1) Models pre-trained on histopathology-specific datasets consistently outperform their ImageNet-pretrained counterparts, evaluated across an average of six tasks. Specifically, the top-performing model achieves an Area Under the Curve (AUC) of 94.72\% on the REC task and 81.23\% on the TD task, with all other tasks surpassing an AUC of 70\%.
(2) Stain normalization leads to a significant decrease in AUC for the prognosis task (REC), while it notably improves AUC for five diagnostic tasks, suggesting REC's reliance on original color properties. 
(3) Within the multi-task learning framework, GradNorm~\citep{chen2018gradnorm} achieves the highest average AUC, with stain normalization providing an additional performance boost. Nevertheless, the multi-task models underperform their single-task counterparts by an average of 3.34\% across the six tasks. This gap underscores the challenge of effectively balancing competing objectives in comprehensive computer-aided diagnosis (CAD) systems.

\section{Related Work}
\label{sec:related_work}

\subsection{Computer-aided Diagnosis and Prognosis for OSCC}

Current research on OSCC often utilizes lesion-focused radiological data and oral photographs, as well as gene and histopathology data from a cellular perspective, to develop models for tumor diagnosis, staging, and prognosis. \cite{ren2020machine} developed a random forest model leveraging radiomic features extracted from head medical images to detect LNM. \cite{fu2020deep} collected 44,409 oral images and developed a deep learning model to classify OSCC images, achieving diagnostic accuracy comparable to that of clinicians.  Using data from TCGA-HNSC~\citep{zuley2016cancer}, \cite{vollmer2024multimodal} developed a random survival forest model that integrates clinical data, genomic profiles, and features extracted from histopathology images to perform survival prediction.  Based on the OSCC dataset~\citep{rahman2020histopathological},  \cite{afify2023novel} developed a ResNet-101 model to classify normal and OSCC images. The ORCHID dataset~\citep{chaudhary2024high} provides a benchmark for analysis, utilizing a DCNN model to classify histopathology images by first categorizing them into normal, OSCC, and OSMF, and then further classifying the OSCC images based on TD labels. Additionally, \cite{zhou2024pathology} developed a deep learning model with semi-supervised learning that utilizes histopathology images to identify critical prognostic factors. These OSCC datasets are often restricted from public access, limited in the patient cohort size, or only considered for diagnosis or prognosis tasks, making them less comprehensive.

\subsection{Histopathology Image Analysis Algorithms}

In the analysis of disease diagnosis and postoperative prognosis for patients with OSCC, histopathology images are considered a powerful foundation by researchers at the hospital. To model patients' histopathology images, vision algorithms are typically used to extract high-dimensional features, which are then employed for specific tasks.
In traditional machine learning, tools like CellProfiler~\citep{stirling2021cellprofiler} are used to extract handcrafted morphological and spatial features from region of interests (ROIs) on histopathology images. Additionally, researchers like~\cite{corredor2019spatial} have designed custom features, employing watershed segmentation and graph theory to analyze tumor-infiltrating lymphocytes for recurrence prediction.
Deep learning has emerged as a powerful tool for extracting features from images. For whole-slide images (WSIs), recent studies have employed multiple instance learning (MIL) algorithms to modeling large-scale images \citep{chen2022scaling, yan2024shapley}. For high-resolution microscope images, researchers resize the images to dimensions suitable for common vision models or apply cropping techniques \citep{chaudhary2024high, albalawi2024oral}. With the advancement of high-throughput data, some institutions have acquired large datasets of histopathology images and trained powerful backbone networks, such as PathoBench~\citep{kang2023benchmarking}, Hibou~\citep{nechaev2024hibou}, CONCH~\citep{lu2024visual}.  These developments have provided valuable support for our research.

\section{Dataset}
\label{sec:dataset}

\subsection{Multi-OSCC Dataset}

We recruited patients diagnosed with OSCC at Sun Yat-sen Memorial Hospital, Sun Yat-sen University, between 2015 and 2022 who had undergone surgical treatment. The inclusion criteria are: (1) a confirmed pathological diagnosis of squamous cell carcinoma; (2) receipt of surgical intervention; and (3) participation in follow-up care for at least two years post-surgery. The study is approved by the Ethics Committee of Sun Yat-sen Memorial Hospital, Sun Yat-sen University, for the use of identifiable human materials and data. Approval is granted on the condition that the study did not involve personal privacy or commercial interests, and consent for exemption from informed consent is obtained.

During the histopathology sections preparation process, the tumor lesions are fixed with formalin solution, dehydrated, embedded in paraffin, sliced, stained by hematoxylin-eosin (H\&E), and sealed to make histopathology slides. After the diagnosis by the pathologists, according to the level of cell differentiation of the tumor, epithelial tissue arrangement, cancer cell nests, histopathology mitotic images, \emph{etc.}, the core and borderline sites of the lesion are selected and the histopathology pictures are preserved after magnification by Olympus microscope. Thus, each patient has two tissue sections taken, one from the core of the lesion and the other from the boundary of the lesion. 

\begin{figure}[!t]
    \centerline{\includegraphics[width=\columnwidth]{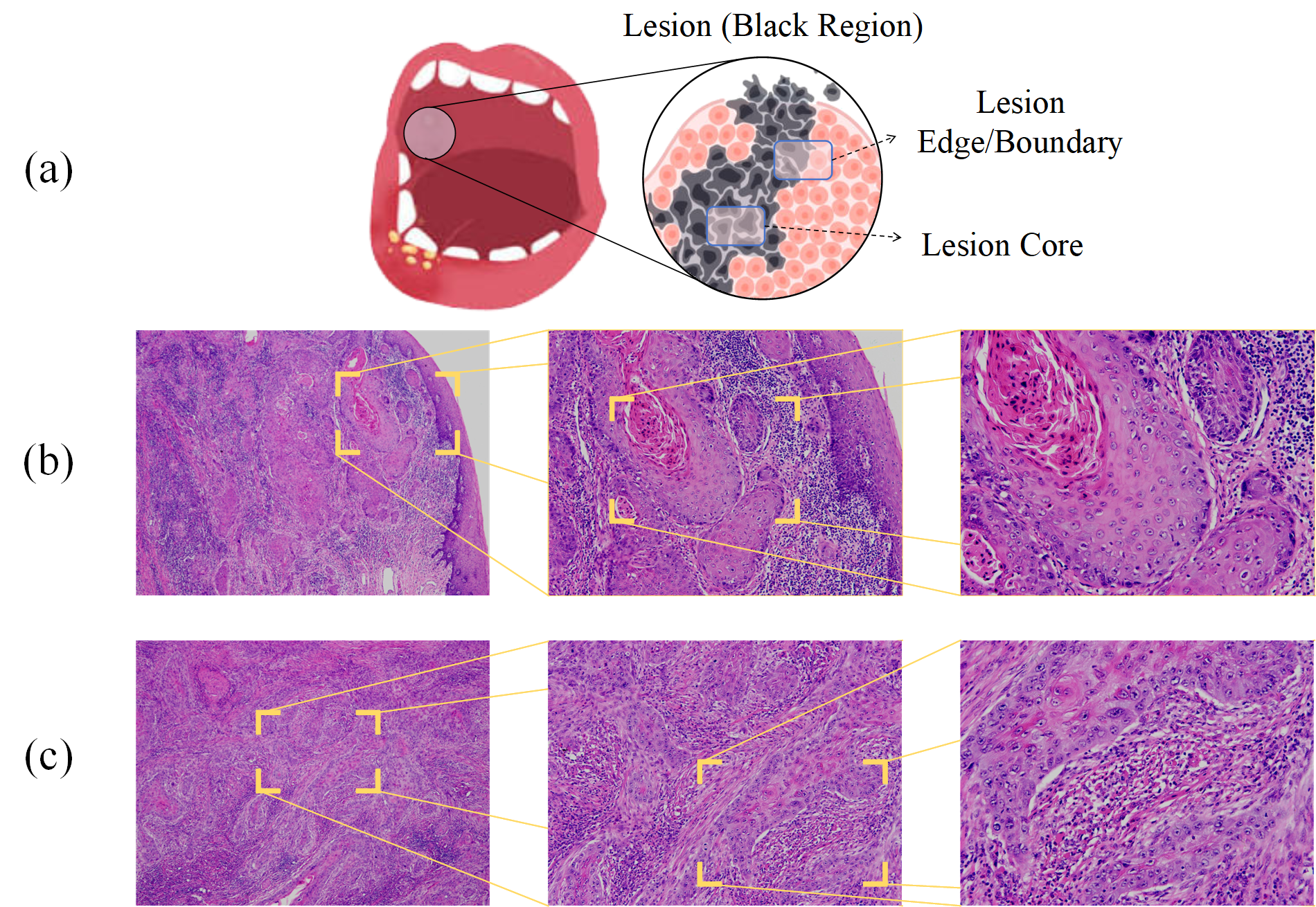}}
    \caption{(a) illustrates an abstract scene depicting the collection process of core/edge histopathology slides. Six images in (b,c) are from the same patient. The three in (b) are captured from tissue sections at the lesion core, with magnifications of \(\times 200\), \(\times 400\), and \(\times 1000\) from left to right, while the images in (c) are from the lesion boundary. In (b), the focus is on keratin pearl details and surrounding cells, while (c) emphasizes cancerous tissue and nearby structures as magnification increases.}
    \label{fig:image_example}
\end{figure}

As noted by \cite{chen2022scaling}, high-magnification images capture details of individual cells and fine-grained features, such as stroma, tumor cells, and lymphocytes.  Mid-magnification images emphasize local clusters of cell-to-cell interactions, highlighting tumor cellularity.  In contrast, low-magnification images provide a global perspective on the interactions and spatial organization of cell clusters within the tissue, including insights into tumor-immune localization. Another study~\citep{lu2021data}, which analyzed image data captured using microscopes and cellphones, highlights that in resource-limited settings, images are often acquired using basic equipment, such as microscopes, rather than advanced scanners.

Building on these insights, we collected histopathology images of each tissue section in our dataset using an optical microscope at magnifications of \(\times 200\), \(\times 400\), and \(\times 1000\) (with a \(\times 10\) eyepiece lens). Each patient is represented by six images in total, with two images per magnification level, each from a different tissue section, and each image has a resolution of \(2592 \times 1944\) pixels. Specific examples are illustrated in Figure~\ref{fig:image_example}. The pathologists ensured that the collected histopathology images encompass the most critical structures of interest, including cancer cells, cancer nests, keratin pearls, nuclear atypia, and necrotic areas, among others. However, there is no guarantee that structures such as nerve fibers and blood vessels are present in every image, due to the challenges associated with slide examination. Patient-level annotations, including diagnosis and prognosis, are obtained from the hospital's electronic medical records.

\subsection{Statistic Analysis}

In this study, we use the Spearman correlation coefficient~\citep{spearman1961proof} to analyze the correlations between tasks, with the correlation coefficients ($r$-value) and $p$-value shown in Table~\ref{tab:multitask_correlation}. A higher correlation coefficient indicates a stronger relationship between tasks, while a $p$-value below 0.05 suggests statistical significance rather than random chance. Except for the task pair (REC, LNM), all inter-task comparisons yielded $p$-values less than 0.05, indicating statistically significant correlations between the analyzed tasks.

Clinically, the six proposed tasks are positively correlated, as more aggressive tumors tend to invade surrounding tissues, increasing the likelihood of lymphatic, neural, and vascular invasion while elevating recurrence risks. Consequently, all correlation coefficients are greater than 0.

A notable observation from the label distribution analysis is that the prognosis task (REC) shows weaker correlations with diagnostic tasks, while the diagnostic tasks exhibit stronger inter-correlations. This difference can be attributed to the nature of tumor recurrence, which is a longer-term process influenced by factors beyond tumor severity, such as treatment efficacy, follow-up plans, and the patient's living environment. In contrast, the diagnostic tasks are more closely related, as they are predominantly affected by tumor severity, a shared and interpretable influencing factor. TI and PI exhibit the highest correlation of 0.48, likely because both tasks assess tumor invasion into specific tissues. Clinical experts attribute this stronger association to the dense distribution of nerves in the maxillofacial region, where tumors invading surrounding tissues are more prone to involve nerves than lymph nodes or blood vessels.

\definecolor{RedOrange}{RGB}{255, 100, 0} 
\definecolor{YellowOrange}{RGB}{255, 170, 0} 
\definecolor{LightBlue}{RGB}{173, 216, 230} 

\renewcommand{\arraystretch}{2}
\begin{table}[th] 
  \centering
  \caption{Spearman correlation coefficients with P-values (in parentheses), colors indicate correlation strength (from light blue, yellow, orange to red), and only the lower triangular matrix is shown due to symmetry.}
  \setlength{\tabcolsep}{1.5mm}{
    \begin{tabular}{c|cccccc}
      REC & \textcolor{red}{\makecell{1.0\\(0)}} &  &  &  &  \multicolumn{2}{c}{\makecell{$r$-value $\uparrow$ \\ ($p$-value $\downarrow$)}} \\ 
      LNM & \textcolor{gray}{\makecell{0.0455\\(0.0975)}} & \textcolor{red}{\makecell{1.0\\(0)}} &  &  &  &  \\ 
      TD & \textcolor{YellowOrange}{\makecell{0.1081\\(1.3e-4)}} & \textcolor{RedOrange}{\makecell{0.2173\\(1.8e-15)}} & \textcolor{red}{\makecell{1.0\\(0)}} &  &  &  \\ 
      TI & \textcolor{LightBlue}{\makecell{0.0736\\(7.3e-3)}} & \textcolor{YellowOrange}{\makecell{0.1631\\(2.4e-9)}} & \textcolor{YellowOrange}{\makecell{0.1477\\(9.8e-8)}} & \textcolor{red}{\makecell{1.0\\(0)}} &  &  \\ 
      CE & \textcolor{LightBlue}{\makecell{0.0653\\(1.8e-2)}} & \textcolor{YellowOrange}{\makecell{0.1961\\(6e-13)}} & \textcolor{YellowOrange}{\makecell{0.1501\\(1e-7)}} & \textcolor{YellowOrange}{\makecell{0.102\\(2e-4)}} & \textcolor{red}{\makecell{1.0\\(0)}} &  \\ 
      PI & \textcolor{LightBlue}{\makecell{0.0833\\(2.4e-3)}} & \textcolor{YellowOrange}{\makecell{0.1869\\(7e-12)}} & \textcolor{RedOrange}{\makecell{0.2066\\(7e-14)}} & \textcolor{RedOrange}{\makecell{0.4808\\(1e-77)}} & \textcolor{YellowOrange}{\makecell{0.1216\\(9e-6)}} & \textcolor{red}{\makecell{1.0\\(0)}} \\ 
      \hline
      & REC & LNM & TD & TI & CE & PI \\ 
    \end{tabular}
  }
  \label{tab:multitask_correlation}
\end{table}
\renewcommand{\arraystretch}{1}

The dataset is divided into training, validation, and test sets. Table~\ref{tab:task_data_distribution} presents a detailed distribution of the dataset, showing the distribution of binary and multiclass labels within each task. Initially, all samples are transformed into tuples based on the six labels of tasks~(REC, LNM, TD, TI, CE, PI). Samples with identical tuples are grouped, and within each group, the samples are randomly split into 3 sets. This process ensures that the distribution of each class within every task remains relatively balanced across the different sets, promoting more consistent model training and evaluation.

\begin{table}[htbp]
  \centering
  \caption{Detailed data distribution across training, validation, and test sets for six tasks, including class-specific count.}
    
  \begin{tabular}{ccccc}
    \toprule
    \multicolumn{2}{c}{Task} & Train & Valid & Test \\
    \midrule
    & & 925 & 200 & 200 \\
    \midrule
    \multirow{2}{*}{Recurrence} & 0 & 745 & 154 & 151 \\
    & 1 & 180 & 46 & 49 \\
    \midrule
    \multirow{2}{*}{\makecell[c]{Lymph Node\\Metastasis}} & 0 & 592 & 119 & 117 \\
    & 1 & 333 & 81 & 83 \\
    \midrule
    \multirow{3}{*}{\makecell[c]{Tumor\\Differentiation}} & 0 & 310 & 68 & 72 \\
    & 1 & 498 & 99 & 95 \\
    & 2 & 117 & 33 & 33 \\
    \midrule
    \multirow{2}{*}{Tumor Invasion} & 0 & 511 & 103 & 101 \\
    & 1 & 414 & 97 & 99 \\
    \midrule
    \multirow{2}{*}{Cancer Embolus} & 0 & 859 & 178 & 176 \\
    & 1 & 66 & 22 & 24 \\
    \midrule
    \multirow{2}{*}{Perineural Invasion} & 0 & 766 & 159 & 156 \\
    & 1 & 159 & 41 & 44 \\
    \bottomrule
    \end{tabular}%
  \label{tab:task_data_distribution}%
\end{table}%

\section{Method}
\label{sec:method}

\begin{figure*}[!t]
    \centering
    \includegraphics[width=0.95\textwidth]{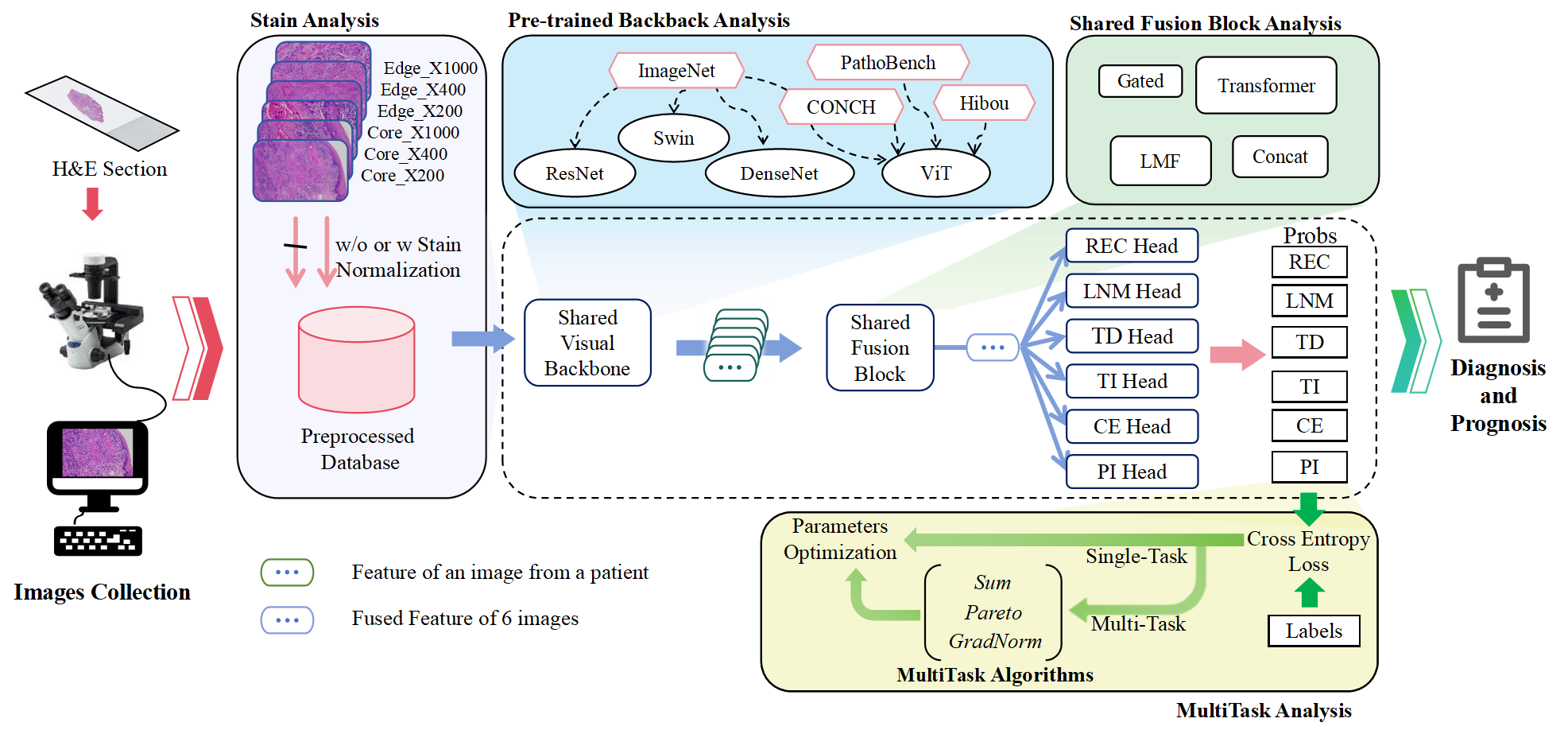}
    \caption{Overview of the proposed pipeline, which includes stain normalization, pre-trained backbone analysis, feature fusion, and multi-task optimization. The pipeline processes six input images from a patient, outputting the probability for a single task in single-task mode or for all tasks in multi-task mode.}
    \label{fig:model_architecture}
\end{figure*}

\subsection{Data Preprocessing}

\subsubsection{Stain Normalization}

In examining the effect of stain inconsistency on diagnostic and prognostic tasks within our dataset, we apply three well-established stain normalization techniques: Reinhard~\citep{reinhard2001color}, Vahadane~\citep{vahadane2016structure}, and Macenko~\citep{macenko2009method}. For each image in the training set, a random image from the same set is chosen as a reference, and the corresponding stain normalization method is used to align the stain profile of the original image with that of the target. This approach results in four distinct training datasets: the original dataset and three versions augmented by different normalization methods. The final performance is evaluated on each dataset, allowing us to analyze the impact of various stain normalization techniques on generalization across different stain variations.

\subsubsection{Images Transform}

Several data augmentation techniques are implemented during model training to enhance the diversity and robustness of the training data. Due to the large size of the original images, we resize all images to $512 \times 512$ pixels. For data augmentation, we always apply z-score normalization, random cropping, and random rotation, while other techniques such as contrast adjustment, sharpness adjustment, horizontal/vertical flipping, and contrast adjustments are applied with a 50\% probability. For the validation and test sets, normalization alone is applied to maintain consistency in evaluation. Furthermore, the Synthetic Minority Oversampling Technique (SMOTE) \citep{chawla2002smote} is used to address class imbalance, generating synthetic samples for underrepresented classes.

\subsection{Model Architecture}

Compared to other microscope image datasets, the challenges in our Multi-OSCC dataset include fusing features from multiple images at varying magnifications and handling multiple tasks. To address these, we design a network architecture for multi-image feature fusion, enabling flexible integration of pre-trained models, feature fusion modules, and multi-task learning optimization algorithms. The architecture of our network is shown in Figure~\ref{fig:model_architecture}.

\subsubsection{Vision Backbone}    

We explore several popular vision models in recent years, including ResNet~\citep{he2016deep}, DenseNet~\citep{huang2017densely}, Vision Transformer~(ViT)~\citep{dosovitskiy2020image}, and Swin Transformer V2 (Swin)~\citep{liu2022swin}. These models include both CNN and Transformer architectures. 

\begin{itemize}
    \item [1.] ResNet50 utilizes residual learning through skip connections, containing 50 convolutional layers grouped into residual blocks. 
    \item [2.] DenseNet121 comprises multiple dense blocks, where each layer is directly connected to all subsequent layers through dense connectivity. 
    \item [3.] ViT-Base/Small: Transformer-based models that treat image patches as tokens and leverage self-attention for image classification. Both ViT-Base and ViT-Small have 12 layers, but ViT-Small features a smaller embedding size compared to ViT-Base.
    \item [4.] Swin-Base utilizes hierarchical Transformer structures with shifted-window attention mechanisms for efficient vision tasks. We used the base model which has 24 layers and outputs embeddings of size 1024.
\end{itemize}

Instead of training the models from scratch, we use transfer learning by loading pre-trained models from existing work and fine-tuning the backbone models with end-to-end training. We test seven different backbones, including ResNet50, DenseNet121, ViT-Base, and Swin-Base with ImageNet~\citep{deng2009imagenet} pre-trained weights, ViT-Small with PathoBench~\citep{kang2023benchmarking} weights, and ViT-Base with weights from Hibou~\citep{nechaev2024hibou} and CONCH~\citep{lu2024visual}. We utilize the ImageNet pre-trained model from the \href{https://github.com/huggingface/pytorch-image-models}{timm} library. The detailed pathological pre-trained models are described below.

\begin{itemize}
    \item [1.] PahtoBench: It utilized the DINO~\citep{caron2021emerging} pre-training method to train a ViT-Small model, leveraging a dataset comprising 32.6 million patches from various cancers, all stained with H\&E at two different magnification levels \(\times 20\) and \(\times 40\).
    \item [2.] CONCH: It employs a visual-language model with the ViT-Base architecture as the image encoder. The image encoder is first pre-trained with the iBOT~\citep{zhou2021ibot} method on a dataset of 16 million image patches, covering over 350 cancer subtypes. It is then fine-tuned on a dataset of more than 1.17 million image-caption pairs. The image encoder from the CONCH model serves as the pre-trained weights for this process.
    \item[3.] Hibou-B: The Hibou-B model is pre-trained using 512 million clean patches with the DINOv2~\citep{oquab2023dinov2} pre-training method. The dataset consisted of H\&E and non-H\&E stained slides, human tissues, veterinary biopsies, and is enriched with cytology slides.
\end{itemize}

\subsubsection{Feature Fusion Module} 

We experiment with four different feature fusion modules: Concatenation, Low-rank Multimodal Fusion (LMF)~\citep{liu2018efficient}, Gated Fusion, and Transformer. These modules are designed to merge multiple features into a single representation to help with the subsequent classification tasks. Concatenation simply combines the features from different images. The LMF method, which builds on TensorFusion, enhances computational efficiency by parallelizing the decomposition of tensors and weights using low-rank factors specific to each modality. The Gated Fusion method employs a gating mechanism to regulate the flow of information between features. In our implementation of the gating mechanism, we assume the features of each image are represented as \( e_i \). The process is as follows: \( e_i \) is passed through a sigmoid function to obtain \( a_i \), and through a tanh function to produce \( t_i \). The final output is computed as \( Z = \sum_{i=0}^5 (e_i \cdot t_i) \).  
The Transformer method in this research employs a 2-layer Transformer encoder~\citep{vaswani2017attention} to facilitate information interaction among six feature vectors extracted from six images. This approach ultimately aggregates multi-view representations by computing the mean of enhanced feature embeddings derived from the six images.

These modules are placed after the shared vision backbone to fuse the features extracted from six images of a patient.

\subsubsection{Multi-task Learning Module}

In the context of multi-task learning for various classification tasks, the objective function is defined as
\begin{equation}
    \theta_s^*, \{\theta_t^*\}_{t \in \mathcal{T}} = \mathop{argmin}_{\theta_s, \{\theta_t\}_{t \in \mathcal{T}}} \sum_{i=1}^{N} \sum_{t \in \mathcal{T}} \lambda_t \mathcal{L}_t(y_i^t, f(x_i; \theta_s, \theta_t))
    \label{eq:multitask_target}
\end{equation}
$\theta_s$ represents the shared parameters across all tasks, while $\{\theta_t\}_{t \in \mathcal{T}}$ refers to the task-specific parameters for each task $t$, and $\mathcal{T} = \{REC, LNM, TD, TI, CE, PI\}$. The function $f(x_i; \theta_s, \theta_t)$ denotes the model's output for input images $x_i$ using both shared and task-specific parameters. \(\lambda_t\) is a weighting factor for $t$-th task's contribution to the total loss, \(N\) is the total number of samples, \(y_i^t\) is the $t$-th task's ground truth label for $i$-th sample, and \(\mathcal{L}_t(y_i^t, f(x_i; \theta_s, \theta_t))\) represents the loss function for $t$-th task. 

Classic multi-task learning models are generally categorized into three types: hard parameter sharing, soft parameter sharing, or a combination of both \citep{crawshaw2020multi}. In our study, we employ the hard parameter sharing approach due to its efficiency and fewer parameter requirements.Specifically, we evaluate three different multi-task learning algorithms: Sum Loss, GradNorm~\citep{chen2018gradnorm}, and Pareto~\citep{sener2018multi}. The Sum Loss method directly sums the loss of all tasks, while GradNorm dynamically balances the gradient magnitudes across tasks to balance adaptive loss. The Pareto method optimizes multiple objectives by finding solutions that are not dominated by any other, to achieve the pareto optimality. These algorithms are used to optimize the shared and task-specific parameters in our model.

\subsubsection{Classification Head}

For each task, we use a standard multi-layer perceptron (MLP) as the classification head. This MLP architecture includes five layers in total, with four hidden layers sized at 768, 256, 128, and 64 units and a classification layer sized at 2 units. After each hidden layer, we employ the rectified linear unit (ReLU) activation function \citep{glorot2011deep} and layer normalization (LayerNorm) \citep{ba2016layer} to enhance the stability and convergence of the model. To mitigate overfitting, we also apply a dropout \citep{srivastava2014dropout} with a probability of 0.5 following the final hidden layer, preceding the output layer that maps to the number of classes.

\section{Experiments}
\label{sec:experiments}

In this section, we present the implementation details and results of our analysis. Finally, we provide the benchmark results of our dataset based on the analysis experiments.

\subsection{Implementation Details} 

In the experiments, we use cross-entropy loss as the target function for each task. During fine-tuning, the learning rate for the backbone is set to a lower value of $5 \times 10^{-7}$, while the learning rate for the other parameters is set to $1 \times 10^{-6}$, with a batch size of 16. We use the AdamW optimizer with a weight decay of $6 \times 10^{-5}$ and adjust the learning rate using a cosine annealing scheduler. The models are trained for over 400 epochs until it is converged.

For evaluation, we use five metrics: accuracy (Acc), area under the receiver operating characteristic curve (AUC), F1 score, recall, and precision. Although the final benchmark reports all metrics, the AUC is the primary metric used to select the best-performing model during the analysis phase. When conducting analysis experiments, it is important to use statistical estimation to assess the generalization performance of a model~\citep{claridge2016estimation}. Thus, we use bootstrap estimation~\citep{diciccio1996bootstrap} to calculate the confidence intervals 95\% for the metrics, providing more detailed model results. All models are trained on a GeForce RTX 3090 (24GB) with fixed random seeds to ensure reproducibility.

\subsection{Backbone Model Analysis}

\begin{table*}[ht]
    \centering
    \caption{Test AUC results of different visual encoders. Bold numerals indicate the highest metric for each task, underlined numerals denote the second-highest metric, and each metric is accompanied by its 95\% confidence interval in parentheses.}
\adjustbox{width=\textwidth}{
    \begin{tabular}{ccccccccr}
      \toprule
      \multirow{2}[2]{*}{\textbf{\makecell{Model \\ (Params)}}} & \multirow{2}[2]{*}{\textbf{Pretraining}} & \multicolumn{7}{c}{\textbf{Test AUC (\%)}} \\
      \cmidrule{3-9}    
      &       & \textbf{REC} & \textbf{LNM} & \textbf{TD} & \textbf{TI} & \textbf{CE} & \textbf{PI} & \multicolumn{1}{c}{\textbf{Mean}} \\
      \midrule
      \vspace{0.1cm} 
      
      \makecell{ResNet50\\(89.68MB)} & Imagenet & \makecell{66.12 \\ (54.63, 72.69)} & \makecell{67.41 \\ (59.54, 74.48)} & \makecell{61.23 \\ (51.47, 68.85)} & \makecell{62.97 \\ (54.49, 68.42)} & \makecell{60.23 \\ (49.78, 72.42)} & \makecell{63.18 \\ (50.18, 68.85)} & 63.52 \\
      \vspace{0.1cm} 
      
      \makecell{DenseNet121\\(26.53MB)} & Imagenet & \makecell{55.18 \\ (45.89, 63.23)} & \makecell{55.03 \\ (45.54, 62.58)} & \makecell{66.4 \\ (56.62, 72.8)} & \makecell{66.36 \\ (58.49, 72.73)} & \makecell{61.13 \\ (47.88, 67.39)} & \makecell{66.71 \\ (57.5, 74.36)} & 61.80 \\
      \vspace{0.1cm} 
      
      \makecell{Swin-Base\\(331.47MB)} & Imagenet & \makecell{85.78 \\ (79.41, 90.65)} & \makecell{66.85 \\ (59.13, 73.3)} & \makecell{76.22 \\ (69.25, 83.46)} & \makecell{\underline{69.37} \\ (62.52, 75.11)} &  \makecell{\textbf{74.67} \\ (61.47, 81.8)} & \makecell{\underline{68.66} \\ (58.49, 75.16)} & 73.59 \\
      \vspace{0.1cm} 
      
      \makecell{ViT-Base\\(344.14MB)} & Imagenet & \makecell{92.13 \\ (88.34, 95.43)} & \makecell{64.31 \\ (57.01, 72.34)} & \makecell{70.94 \\ (62.22, 79.11)} & \makecell{67.07 \\ (59.55, 74.55)} & \makecell{70.80 \\ (60.23, 81.65)} & \makecell{69.90 \\ (61.02, 78.09)} & 72.52 \\
      \vspace{0.1cm} 
      
      \makecell{ViT-Small\\(83.86MB)} & Imagenet & \makecell{91.33 \\ (85.89, 95.46)}  &  \makecell{63.43 \\ (55.17, 71.08)} & \makecell{75.99\\ (68.14, 82.95)} &  \makecell{63.62 \\ (55.81, 71.24)} &   \makecell{58.21 \\  (45.85, 71.20)}  & \makecell{63.73 \\ (53.74, 73.88)} & 69.39  \\
      \vspace{0.1cm} 
      
      \makecell{ViT-Small\\(75.54MB)} & PathoBench & \makecell{\underline{93.13} \\ (89.18, 96.47)} &  \makecell{69.35 \\ (62.04, 76.59)} & \makecell{75.88 \\ (68.19, 82.87)} &  \makecell{\textbf{72.31} \\ (64.94, 79.35)} &\makecell{\underline{73.18} \\ (62.20, 84.34)}  & \makecell{66.14 \\ (57.02, 75.39)} & \textbf{75.00}  \\
      \vspace{0.1cm} 
      
      \makecell{ViT-Base\\(344.82MB)} & CONCH & \makecell{80.7 \\ (72.43, 86.98)} & \makecell{\textbf{70.29} \\ (61.9, 75.68)} & \makecell{\textbf{81.23} \\ (74.15, 87.58)} & \makecell{67.45 \\ (61.47, 73.76)} & \makecell{68.16 \\ (54.98, 76.89)} & \makecell{\textbf{74.17} \\ (63.52, 79.77)} & 73.67 \\
      \vspace{0.1cm} 
      
      \makecell{ViT-Base\\(327.60MB)} & Hibou-B & \makecell{\textbf{94.72} \\ (89.78, 97.35)} & \makecell{\underline{69.39} \\ (61.44, 75.03)} & \makecell{\underline{78.36} \\ (70.79, 84.64)} & \makecell{64.22 \\ (55.68, 70.46)} & \makecell{73.08 \\ (59.36, 79.02)} & \makecell{67.77 \\ (57.05, 75.69)} & \underline{74.59} \\
      \bottomrule
    \end{tabular}%
    }

    \label{tab:backbone_table}
\end{table*}

Different backbones perform differently on various datasets. In this experiment, we select the simplest fusion method, Concat, to combine the features extracted from multiple images and analyze the performance of different backbone models in our data set, with the results shown in Table~\ref{tab:backbone_table}. Across all tasks, the histopathology-specific pre-training achieves a higher average AUC than models initialized with ImageNet pre-trained weights.

Besides, the ViT-Small model with PathoBench pre-trained weights achieves the highest average AUC of 75.00\%, while the ViT-Base model with Hibou-B pre-trained weights ranks second with an average AUC of 74.59\%. The ViT-Base model with CONCH pre-trained weights shows weaker average performance but still achieves top-1 AUC in three individual tasks, demonstrating strong generalization capability. The results indicate that different histopathology-specific pre-training strategies yield varying results.

Although Hibou-B falls short in the overall average comparison, it outperforms the PathoBench model in four tasks (REC, LNM, TD, PI), which can be attributed to its larger pre-training dataset and more diverse data sources. The CONCH model, pre-trained on a smaller dataset, still benefits from the image-caption pairs, which may contribute to its improved performance in certain tasks.


Given the highest average AUC of the PathoBench pre-trained ViT-Small, combined with its relatively low number of parameters, we select it as the base model for subsequent analysis experiments.

\begin{table}[htbp]
    \centering
    \caption{Test AUC for different feature fusion methods in feature fusion analysis.}
    \adjustbox{width=\columnwidth}{
    \begin{tabular}{lcccc}
        \toprule
        & \multicolumn{4}{c}{\textbf{Test AUC (\%)}} \\ 
        \multirow{2}{*}{\textbf{Task}} & \multicolumn{4}{c}{\textbf{Fusion Block (Params)}} \\ 
        \cmidrule{2-5}
        & \textbf{\makecell{Concat\\(6.76MB)}} & \textbf{\makecell{Transformer \\  (37.19MB)}} & \textbf{\makecell{LMF\\(75.53MB)}} & \textbf{\makecell{Gated\\(1.13MB)}} \\ 
        \midrule
        REC   & \makecell{\textbf{93.13} \\ (89.18, 96.47)} & \makecell{\underline{90.71} \\ (85.67, 94.89)} & \makecell{86.86 \\ (80.5, 92.31)} & \makecell{90.85 \\ (85.26, 94.54)} \\
        LNM   & \makecell{\textbf{69.35} \\ (62.04, 76.59)} & \makecell{\underline{68.86} \\ (60.5, 74.54)}& \makecell{64.95 \\ (56.33, 70.65)} & \makecell{67.71 \\ (59.23, 73.62)} \\
        TD    & \makecell{75.88 \\ (68.19, 82.87)} & \makecell{\textbf{77.59} \\ (69.54, 84.52)} & \makecell{\underline{77.43} \\ (69.48, 83.87)} & \makecell{76.36 \\ (68.2, 83.01)} \\
        TI    & \makecell{\textbf{72.31} \\ (64.94, 79.35)}& \makecell{64.76 \\ (57.58, 71.59)} & \makecell{\underline{70.54} \\ (60.37, 73.98)} & \makecell{63.99 \\ (56.5, 71.23)} \\
        CE    & \makecell{\textbf{73.18} \\ (62.20, 84.34)} & \makecell{\underline{70.95} \\ (58.03, 79.99)} & \makecell{68.13 \\ (57.41, 77.59)} & \makecell{63.73 \\ (51.24, 73.64)}\\
        PI    & \makecell{66.14 \\ (57.02, 75.39)} & \makecell{\underline{68.29} \\ (57.93, 73.04)} & \makecell{\textbf{69.31} \\ (57.46, 72.79)} & \makecell{64.33 \\ (50.76, 67.93)} \\
        \midrule
        Mean  & \textbf{75.00} & \underline{73.41} & 72.87 & 71.11  \\
        \bottomrule
    \end{tabular}}
    \label{tab:feature_fusion_table}
\end{table}

\begin{table}[!t]
      \centering
      \caption{Test AUC for Core and Edge Regions in Multi-slice Analysis.}

      \begin{tabular}{c c c c}
        \toprule
        \multirow{2}{*}{\textbf{Task}} & \multicolumn{3}{c}{\textbf{Test AUC (\%)}} \\ 
        \cmidrule{2-4}
        & \textbf{Core} & \textbf{Edge} & \textbf{Core + Edge} \\ 
        \midrule
        \vspace{0.1cm} 
        REC  & \makecell{86.47 \\ (78.84, 92.06)} & \makecell{\underline{91.01} \\ (86.07, 94.36)} & \makecell{\textbf{93.13} \\ (89.18, 96.47)} \\
        \vspace{0.1cm} 
        LNM  & \makecell{61.57 \\ (49.63, 65.6)} & \makecell{\textbf{71.68} \\ (59.22, 74.42)} & \makecell{\underline{69.35} \\ (62.04, 76.59)} \\
        \vspace{0.1cm} 
        TD   & \makecell{\underline{73.16} \\ (60.27, 75.98)} & \makecell{72.57 \\ (63.33, 79.16)} & \makecell{\textbf{75.88} \\ (68.19, 82.87)} \\
        \vspace{0.1cm} 
        TI   & \makecell{67.16 \\ (56.04, 70.54)} & \makecell{\underline{68.76} \\ (56.49, 70.32)} & \makecell{\textbf{72.31} \\ (64.94, 79.35)} \\
        \vspace{0.1cm} 
        CE   & \makecell{65.62 \\ (53.99, 75.83)} & \makecell{\underline{72.44} \\ (60.75, 83.62)} & \makecell{\textbf{73.18} \\ (62.20, 84.34)} \\
        \vspace{0.1cm} 
        PI   & \makecell{63.56 \\ (52.84, 72.7)} & \makecell{\textbf{69.40} \\ (60.24, 76.88)} & \makecell{\underline{66.14} \\ (57.02, 75.39)} \\
        \midrule
        Mean & 69.59 & \underline{74.73} & \textbf{75.00} \\
        \bottomrule
      \end{tabular}
    
      \label{tab:multi_site_table}
    \end{table}

\begin{table}[!t]
    \centering
    \caption{Test AUC results of stain normalization methods.}
    \adjustbox{width=\columnwidth,center}{ 
    \setlength{\tabcolsep}{1.2mm}{ 
    \begin{tabular}{ccccc}
        \toprule
        \multirow{2}[0]{*}{\textbf{Task}} 
        & \multicolumn{4}{c}{\textbf{Test AUC (\%)}} \\
        \cmidrule{2-5}
        & \textbf{Origin} & \textbf{Reinhard} & \textbf{Vahadane} & \textbf{Macenko} \\
        \midrule
        REC   & \makecell{\textbf{93.13} \\ (89.18, 96.47)} 
              & \makecell{\underline{90.53} \\ (84.51, 92.99)} 
              & \makecell{87.12 \\ (79.97, 90.37)} 
              & \makecell{87.15 \\ (79.7, 91.25)} \\
        LNM   & \makecell{69.35 \\ (62.04, 76.59)} 
              & \makecell{\textbf{71.06} \\ (64.15, 77.92)} 
              & \makecell{69.97 \\ (60.72, 74.53)}
              & \makecell{\underline{70.24} \\ (61.47, 74.85)}\\
        TD    & \makecell{75.88 \\ (68.19, 82.87)} 
              & \makecell{75.21 \\ (67.40, 82.09)} 
              & \makecell{\underline{76.54} \\ (66.63, 83.05)} 
              & \makecell{\textbf{77.37} \\ (67.68, 82.62)} \\
        TI    & \makecell{72.31 \\ (64.94, 79.35)} 
              & \makecell{\underline{72.47} \\ (65.10, 79.54)} 
              & \makecell{69.25 \\ (59.52, 72.98)} 
              & \makecell{\textbf{72.6} \\ (66.2, 77.82)}\\
        CE    & \makecell{73.18 \\ (62.20, 84.34)} 
              & \makecell{\textbf{75.52} \\ (63.41, 86.44)} 
              & \makecell{72.51 \\ (59.37, 81.79)} 
              & \makecell{\underline{74.41} \\ (58.3, 82.05)} \\
        PI    & \makecell{66.14 \\ (57.02, 75.39)} 
              & \makecell{\underline{66.19} \\ (56.52, 75.42)} 
              & \makecell{\textbf{70.73} \\ (59.53, 76.11)}   
              & \makecell{64.11 \\ (54.21, 71.78)} \\
        \midrule
        Mean & \underline{75.00} & \textbf{75.16} & 74.35 & 74.31 \\
        \bottomrule
    \end{tabular}}}
    \label{tab:stain_table}
\end{table}

\begin{table*}[htbp]
  \centering
    \caption{Test AUC results of multi-task learning methods.}
  \resizebox{0.80\textwidth}{!}{
    \begin{tabular}{c  c  c c c | c c }
      \toprule
       & \multicolumn{6}{c}{\textbf{Test AUC (\%)}} \\ 
      \textbf{\multirow{2}{*}{\textbf{Task}}} & \multicolumn{4}{c}{\textbf{W/o Stain Normalization}} &  \multicolumn{2}{c}{\textbf{With Stain Normalization (Reinhard)}} \\
      
      \cmidrule(lr){2-5}\cmidrule(lr){6-7}

       & \textbf{Single Task} & \textbf{Sum Loss} & \textbf{GradNorm} & \textbf{Pareto} & \textbf{Single Task} & \textbf{GradNorm} \\ 
      \midrule
      \vspace{0.1cm} 
      REC  & \makecell{\textbf{93.13} \\ (89.18, 96.47)}  & \makecell{83.4 \\ (76.5, 89.2)} & \makecell{89.78 \\ (84.38, 94.01)} & \makecell{85.16 \\ (78.78, 90.4)} & \makecell{\underline{90.53} \\ (84.51, 92.99)}   & \makecell{87.17 \\  (82.07, 91.73)}  \\ 
      \vspace{0.1cm} 
      LNM  & \makecell{\underline{69.35} \\ (62.04, 76.59)} & \makecell{63.52 \\ (55.65, 70.53)} & \makecell{61.46 \\ (53.66, 69.51)} & \makecell{62.65 \\ (55.02, 68.71)} &  \makecell{\textbf{71.06} \\ (64.15, 77.92)}  & \makecell{67.61 \\    (60.01, 75.41)}   \\
      TD   & \makecell{\textbf{75.88} \\ (68.19, 82.87)} & \makecell{69.28 \\ (64.96, 79.5)} & \makecell{70.27 \\ (61.11, 78.59)} & \makecell{67.76 \\ (59.35, 76.26)} & \makecell{\underline{75.21} \\ (67.40, 82.09)} & \makecell{72.51 \\    (63.87, 80.34)}  \\ 
      \vspace{0.1cm} 
      TI   & \makecell{\underline{72.31} \\ (64.94, 79.35)} & \makecell{66.27 \\ (59.37, 72.98)} & \makecell{69.23 \\ (62.13, 76.40)} & \makecell{57.66 \\ (48.45, 63.89)}  & \makecell{\textbf{72.47} \\ (65.10, 79.54)} & \makecell{67.82 \\    (60.62, 75.27)}   \\ 
      \vspace{0.1cm} 
      CE   & \makecell{73.18 \\ (62.20, 84.34)} & \makecell{65.98 \\ (53.57, 79.89)} & \makecell{62.57 \\ (48.49, 75.77)} & \makecell{\textbf{76.7} \\ (64.16, 83.52)} & \makecell{\underline{75.52} \\ (63.41, 86.44)} & \makecell{68.37 \\   (55.07, 81.15)} \\ 
      \vspace{0.1cm} 
      PI   &  \makecell{66.14 \\ (57.02, 75.39)}  & \makecell{64.87 \\ (55.03, 70.38)} & \makecell{\textbf{69.22} \\ (60.42, 77.95)} & \makecell{65.46 \\ (52.64, 70.97)} & \makecell{66.19 \\ (56.52, 75.42)} & \makecell{\underline{67.41} \\  (58.52, 76.13)} \\ 
      \midrule
      Mean & \underline{75.00} & 68.89 & 70.42 & 69.23 & \textbf{75.16} & 71.82        \\
      \bottomrule
    \end{tabular}%
  }
 
    \label{tab:multi_task_table}
\end{table*}

\begin{figure}[t]
    \centering
    \includegraphics[width=\linewidth]{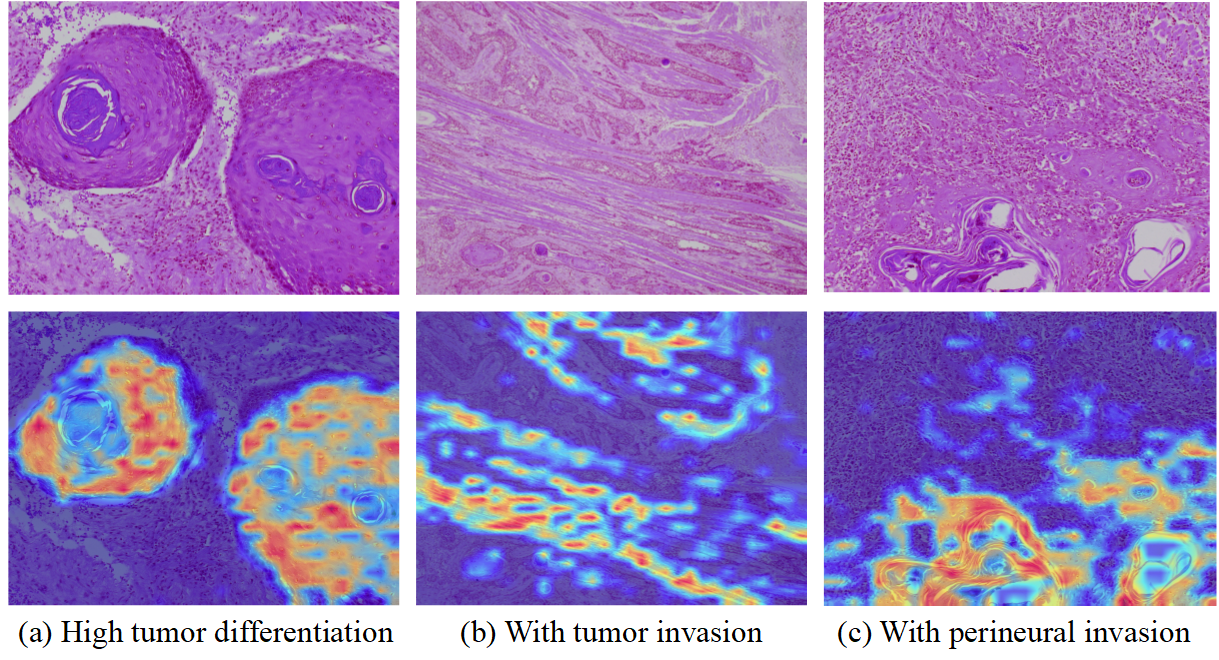}
    \includegraphics[width=\linewidth]{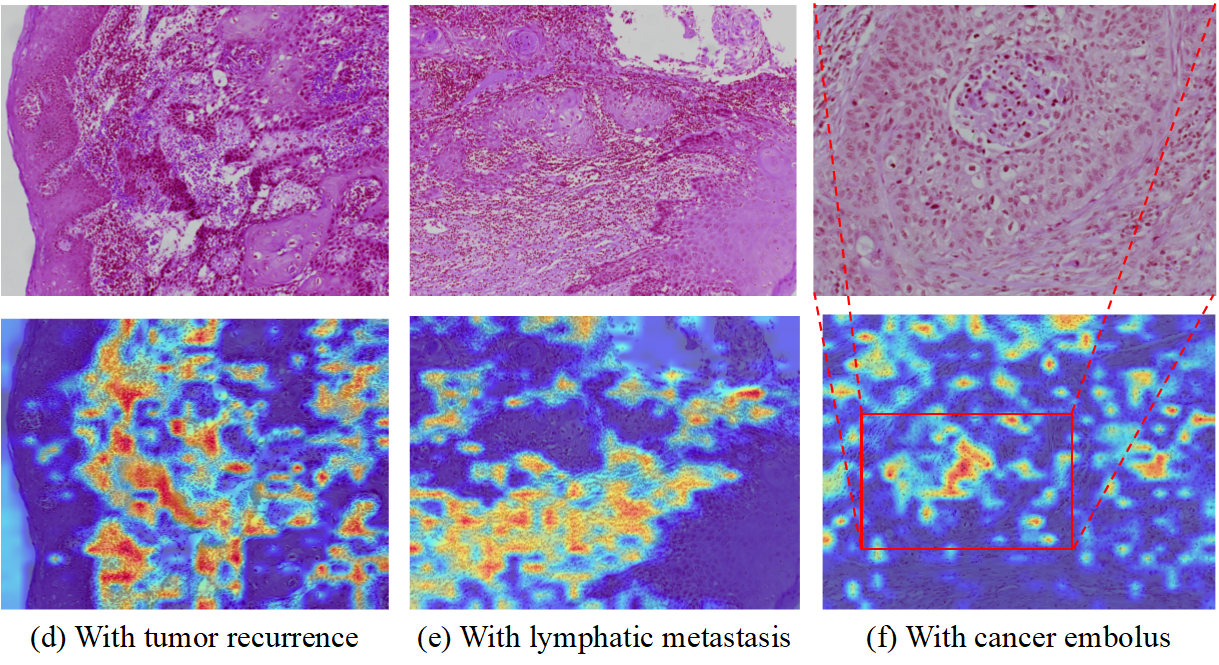}
    \caption{Visualization of model attention for each of the six clinical tasks, generated by the top-performing model for that respective task. In each subfigure, the original histopathology image (top) is paired with a heatmap (bottom) indicating the model's focus areas. (a) High tumor differentiation, with keratin pearls highlighted as clinical evidence of high differentiation.    (b) Tumor invasion into surrounding tissues, showing highlighted regions of infiltrated striated muscle.    (c) Perineural invasion, where the model highlights keratin pearls and the tumor cell nests on the right.    (d) Tumor recurrence and (e) lymph node metastasis, where the model appears to focus on both the tumor regions and surrounding structure.    (f) Cancer emboli, with the embolus location prominently highlighted.}
    \label{fig:results_visualization}
\end{figure}

\subsection{Feature Fusion Analysis}

In our pipeline, we use a post-fusion method to combine features from multiple images of a single patient. However, the choice of fusion method significantly impacts model performance. Therefore, further exploration of different feature fusion techniques for integrating features from multiple histopathology images is essential. The specific results are presented in Table~\ref{tab:feature_fusion_table}. From the AUC results, both Concat and Transformer perform well. Concat achieves the best results in tasks such as REC, LNM, TI, and CE, while Transformer and LMF excel in tasks like TD and PI. However, when averaging the AUC across all tasks, Concat outperforms the others. Given its simplicity and effectiveness, we select Concat for future experiments.

\subsection{Multi-slice Analysis}
\label{subsec:multi_slice_analysis}

Compared to other publicly available histopathology image datasets, our Multi-OSCC dataset presents a challenge by including six histopathology images per patient. In this section, we test the impact of using multiple images on model performance. The results are shown in Table~\ref{tab:multi_site_table}. We set up three groups: using only Core lesion images, only Edge lesion images, and using all images (Core+Edge). The Core+Edge model performs best in tasks such as REC, TD, TI, and CE, and comes second in the LNM and PI tasks. However, the performance improvement for the CE and PI tasks is less pronounced, highlighting the challenge of fusing multiple image features. The Core+Edge model has the highest average AUC at 75.00\%, compared to 69.59\% for Core-only and 74.73\% for Edge-only. This demonstrates that while adding more histopathology images increases the complexity of feature fusion, it also improves the potential for better model performance.

\subsection{Stain Normalization Analysis}

Stain normalization is widely used as an augmentation technique for histopathology image datasets, but in our analysis, it produces inconsistent results. The experimental outcomes are shown in Table~\ref{tab:stain_table}. For the prognosis task REC, all three stain normalization methods lead to a significant drop in performance. However, in the other five tasks, stain normalization improves the AUC results in most cases. We hypothesize that the color of histopathology images is a key factor for the REC task, and thus, the bias introduced by stain normalization may lead to a decrease in REC prediction performance. For the other tasks, however, the effect of stain normalization is positive. This suggests that stain normalization has different impacts on different tasks. Therefore, in the single-task benchmark~\ref{tab:backbone_table}, we present the results without stain normalization for the REC task, while for the other tasks, we present the results with Reinhard stain normalization~\citep{reinhard2001color}.

\subsection{Multi-task Analysis}

We adopt the hard parameter sharing paradigm to build a multi-task learning model and test various optimization algorithms, with the comparison results shown in Table~\ref{tab:multi_task_table}.

Although methods such as GradNorm~\citep{chen2018gradnorm} and Pareto optimization~\citep{sener2018multi} outperform the baseline loss summation, the multi-task model still suffers performance drops in several tasks, particularly REC, LNM, TD, and TI, resulting in an average AUC degradation of 3.34\% across all six tasks. This underscores the difficulty of creating a single, universally effective model. Consistent with our earlier analysis, introducing stain normalization reduces performance on the REC task but yields a net improvement in overall multi-task model accuracy.

Since GradNorm performs well in the multi-task experiment, we select it as the optimization method for our final multi-task learning benchmark.

\subsection{Results Visualization}

We employ GradCAM++~\citep{chattopadhay2018grad} to visualize the areas of focus of the benchmark model on the histopathology images. Subsequently, we invite a pathologist to review a subset of correctly predicted images from the validation and test sets with confidence scores higher than 0.7 to interpret the model's attention. Figure~\ref{fig:results_visualization} presents specific visualization examples along with explanations.

\subsection{Ablation Study of Image Resolution}



\begin{figure*}[!t]
    \centerline{\includegraphics[width=\textwidth]{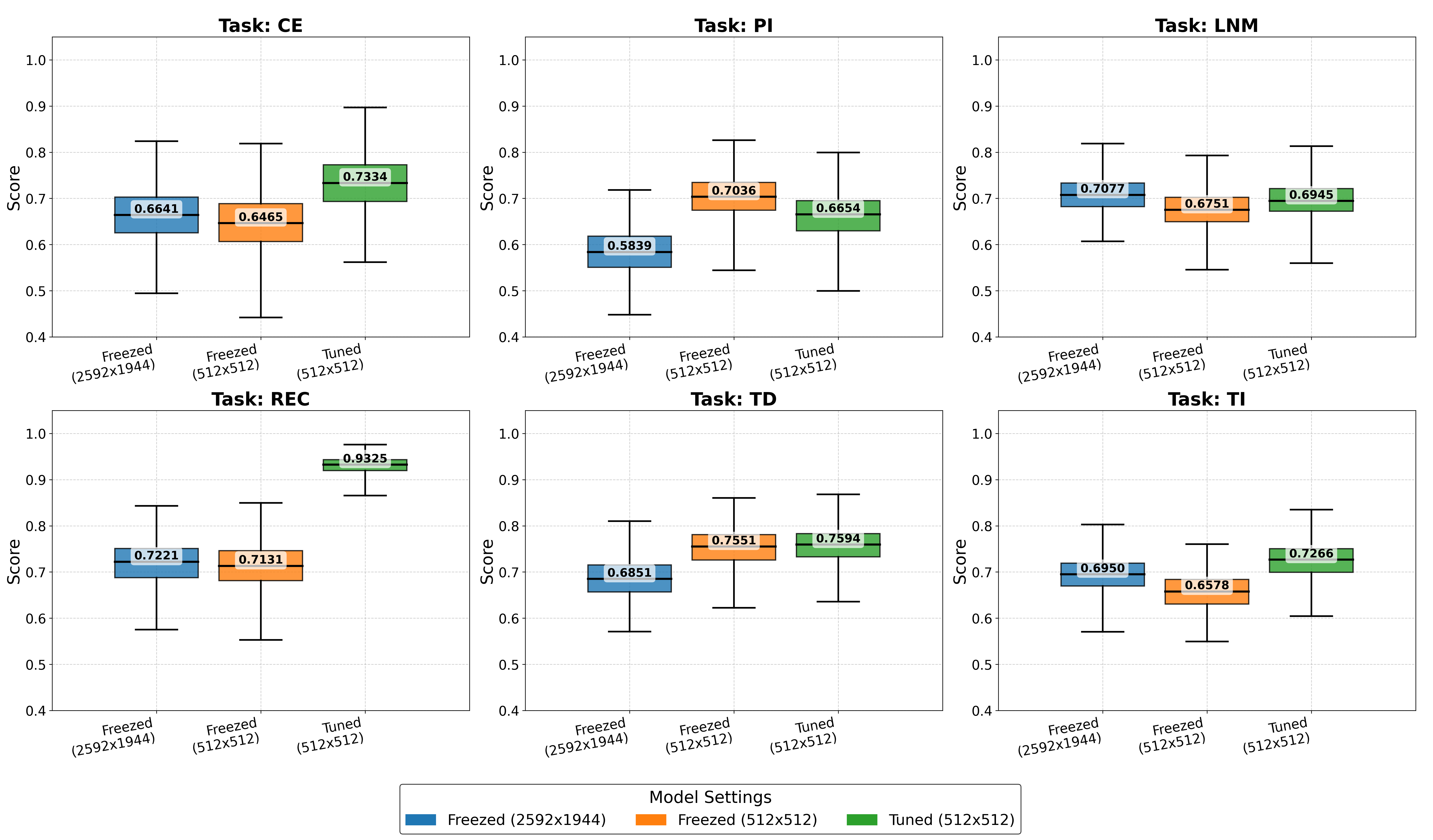}}
    \caption{Performance comparison of \texttt{Freezed (2592x1944)}, \texttt{Freezed (512x512)} and \texttt{Tuned (512x512)} models.}
    \label{fig:further_analysis_figure}
\end{figure*}

The images in our collected dataset possess a high resolution of 2592×1944 pixels. Processing images at this full resolution, particularly when fine-tuning the vision encoder, presents a significant computational challenge. Our estimates indicate that training with full-resolution images would increase the GPU memory consumption by approximately 20-fold compared to using a 512×512 resolution, 
which far exceeds our available hardware resources. To systematically investigate the impact of this resolution reduction on model performance, we conducted a comparative analysis with the following three experimental setups:

\begin{itemize}
    \item  ViT-PathoBench-Freezed (2592×1944): The model utilizes a frozen, pre-trained ViT to extract general visual features from the full-resolution images. In this setting, the encoder's weights are not updated during training.
\item ViT-PathoBench-Freezed (512x512): Similar to the first setup, the ViT encoder is frozen but operates on the downscaled 512x512 images.
\item ViT-PathoBench-Tuned (512x512): The ViT encoder is fine-tuned end-to-end during training using the 512x512 resolution images. This corresponds to our main experimental configuration.
\end{itemize}

This analysis provides a clearer understanding of the trade-offs between image resolution and GPU resources. The comparative results are visualized in Figure \ref{fig:further_analysis_figure}.
The findings reveal several key insights. Firstly, when using a frozen encoder, the full-resolution model (\texttt{Freezed (2592x1944)}) marginally outperforms its downscaled counterpart (\texttt{Freezed (512x512)}) in most tasks, which suggests that some fine-grained details are lost during image resizing.
However, the \texttt{Tuned (512x512)} model consistently and substantially surpasses its \texttt{Freezed (512x512)} counterpart across nearly all tasks, with a particularly notable improvement in Task REC (from $0.7131$ to $0.9325$). This result underscores the paramount importance of fine-tuning the vision encoder, as the generic pathological features learned during pre-training may not be optimal for specialized downstream tasks. Consequently, devising an effective strategy to fine-tune the model using original-resolution images (2592x1944) presents a significant challenge for future research.

\begin{table}[htp]
    \centering
    \caption{Single-task Benchmark. All models adopt ViT-Small + PathoBench as the backbone with concatenation for fusion.  $^\dag$ denotes models trained on original data, while $^*$ denotes models trained with Reinhard stain normalization.}
    \resizebox{\columnwidth}{!}{
      \setlength{\tabcolsep}{1mm}{
      \begin{tabular}{lccccc}
      \toprule
          \multirow{2}{*}{\textbf{Task}} & \multicolumn{5}{c}{\textbf{Test Set Metrics (\%)}} \\
         \cmidrule{2-6}
         & \textbf{Acc}   & \textbf{AUC}   & \textbf{F1}    & \textbf{Recall} & \textbf{Precision} \\
         \midrule
         REC$^\dag$   &   \makecell{87.00 \\ (82.00, 91.00)}    &  \makecell{93.13 \\ (89.18, 96.47)}     &  \makecell{85.19 \\  (78.56, 90.94)}      &    \makecell{81.63 \\ (69.39, 91.84)}   &  \makecell{81.94 \\  (75.83, 86.85)} \\
         \midrule
          LNM$^*$   &    \makecell{64.50 \\ (58.24, 71.00)}  &    \makecell{71.06 \\ (64.15, 77.92)}   &   \makecell{63.38 \\ (56.71, 70.03)}    &   \makecell{63.36 \\ (56.78, 70.19)}    & \makecell{63.40 \\  (56.84, 70.28)}  \\
         TD$^*$   &     \makecell{58.00 \\ (52.50, 64.00)}  &   \makecell{75.21 \\ (67.40, 82.09)}    &  \makecell{57.23 \\  (50.68, 63.59)}    &   \makecell{60.50 \\ (53.49, 66.68)}    & \makecell{57.91 \\ (51.36, 64.72)} \\
         TI$^*$   &  \makecell{69.00 \\ (62.50, 75.00)}     &   \makecell{72.47 \\ (65.10, 79.54)}    &   \makecell{68.29 \\ (61.50, 74.57)}    &  \makecell{69.16 \\  (62.66, 75.14)}     &  \makecell{71.19 \\  (64.33, 77.60)} \\
          CE$^{*}$    &  \makecell{79.50 \\  (73.50, 84.50)}   &  \makecell{75.52 \\ (63.41, 86.44)}     &   \makecell{66.36 \\ (58.97, 73.42)}    &   \makecell{75.76 \\  (65.76, 84.94)}    & \makecell{64.32 \\  (58.73, 69.98)} \\
          PI$^{*}$   &   \makecell{75.50 \\  (70.00, 80.26)}    &   \makecell{66.19 \\ (56.52, 75.42)}    &   \makecell{58.95 \\  (50.88, 66.84)}    &     \makecell{58.19 \\  (51.22, 65.08)}  & \makecell{61.33 \\  (51.76, 70.95)} \\
       \bottomrule
      \end{tabular}%
      }
      }
      \label{tab:single_task_benchmark}%

\end{table}%

\begin{table}[htp]
      \centering
      \caption{Multi-task Benchmark. All models use ViT-Small + PathoBench as the backbone and concatenation for fusion. The dataset was preprocessed with Reinhard stain normalization, and GradNorm was applied as the optimization method.}
      \resizebox{\columnwidth}{!}{
        \setlength{\tabcolsep}{1mm}{
      \begin{tabular}{lccccc}
        \toprule
        \multirow{2}{*}{\textbf{Task}} & \multicolumn{5}{c}{\textbf{Test Set Metrics (\%)}} \\
        \cmidrule{2-6}
         & \textbf{Acc}   & \textbf{AUC}   & \textbf{F1}    & \textbf{Recall} & \textbf{Precision} \\
      \midrule
      REC   &   \makecell{81.00 \\ (76.50, 85.00)}    &   \makecell{87.17 \\ (82.07, 91.73)}    &   \makecell{69.74 \\ (61.39, 76.99)}    &    \makecell{76.00 \\ (66.86, 85.08)}   & \makecell{67.43 \\ (60.33, 74.21)} \\
      LNM   &  \makecell{62.00 \\  (56.00, 68.50)}     &  \makecell{67.61\\ (60.01, 75.41)}     &  \makecell{58.77 \\ (52.04, 65.72)}     &  \makecell{60.24 \\ (53.19, 67.98)}     & \makecell{58.94 \\ (52.71, 65.55)} \\
      TD    &   \makecell{61.50 \\ (55.50, 67.50)}    &    \makecell{72.51 \\ (63.87, 80.34)}   &   \makecell{61.22 \\ (53.88, 67.49)}     &   \makecell{61.44 \\ (54.88, 68.82)}    &  \makecell{61.29 \\  (53.80, 67.71)} \\
      TI    &    \makecell{64.00 \\ (58.00, 70.50)}   &   \makecell{67.82 \\ (60.62, 75.27)}    &  \makecell{63.91 \\ (57.98, 70.44)}     &  \makecell{64.23 \\  (58.45, 70.85)}    &  \makecell{64.06 \\ (58.11, 70.54)} \\
      CE    &   \makecell{86.50 \\ (83.50, 89.00)}    &   \makecell{68.37 \\ (55.07, 81.15)}    &  \makecell{52.79 \\ (45.95, 62.02)}     &  \makecell{58.59 \\  (43.81, 82.14)}     & \makecell{52.74 \\ (48.20, 59.28)} \\
      PI    &   \makecell{76.00 \\ (71.74, 80.26)}    &   \makecell{67.41 \\ (58.52, 76.13)}    &  \makecell{55.36 \\ (47.87, 63.72)}     &  \makecell{60.00 \\ (48.89, 72.39)}     & \makecell{55.24 \\  (49.42, 61.73)}  \\
      \bottomrule
      \end{tabular}%
        }
      }

    \label{tab:multi_task_benchmark}%
\end{table}%

\subsection{Benchmark Results}
\label{subsec:benchmark}
    
    To promote the standardized use of this dataset, we establish a unified benchmark framework. While optimal results for individual tasks may arise from varying configurations, a fair and reproducible benchmark necessitates a consistent setup across all tasks. Through comprehensive experimentation, we have defined our benchmark configuration with the following core components:
    
    \begin{itemize}
        \item \textbf{Backbone}: PathoBench-pretrained ViT-Small
        \item \textbf{Feature Fusion}: Feature concatenation strategy
        \item \textbf{Stain Normalization}: Reinhard method exclusively applied in diagnostic training but excluded from REC tasks due to color sensitive
        \item \textbf{Multi-task Settings}: GradNorm optimization combined with Reinhard stain normalization for image preprocessing
    \end{itemize}

The quantitative results are systematically reported in Table~\ref{tab:single_task_benchmark} (single-task performance) and Table~\ref{tab:multi_task_benchmark} (multi-task performance).

\section{Discussion}
\label{sec:discussion}

In this section, we discuss in more detail the characteristics of our proposed dataset and its value for future research.

Our dataset provides labels for multiple targets, supporting a wide range of studies, including multi-task learning and the development of more generalizable models. In our analysis, the highest AUC for the REC task reached 94.72\%, and the highest AUC for the TD task is 81.23\%. Additionally, the optimal AUC for other tasks exceeded 70\%, demonstrating histopathology images' effectiveness in prognostic predictions and cancer differentiation diagnosis, and enabling research across broader diagnostic applications. 


Our data collection methodology aligns with approaches outlined by \cite{chaudhary2024high} and \cite{rahman2020histopathological}, where histopathology images are captured using a microscope at various high magnifications. We opted for electronic microscopy over WSI for collecting these histopathology images. As detailed in Section~\ref{sec:dataset}, electronic microscopy is a simpler technique. Previous work, such as \cite{lu2021data}, has explored alternative methods for acquiring pathological images (e.g., using mobile phones), which makes this data collection approach feasible in resource-constrained environments. This also addresses the challenge of large data volumes in pathological image analysis from a resolution perspective, as representative regions can be effectively sampled from histopathology images \citep{kayser2009theory}. We acknowledge that relying solely on electronic microscopy, compared to WSI, might lead to some information loss. To mitigate this, we captured images at multiple resolutions and from various lesion locations to preserve more comprehensive information. The high AUC achieved in our benchmark results (Section~\ref{subsec:benchmark}) and the multi-site analysis (Section~\ref{subsec:multi_slice_analysis}) collectively demonstrate the efficacy of supplementing representative histopathology images, thereby validating the feasibility of our data collection method. Therefore, from both a technical standpoint and an analysis of experimental metrics, this dataset possesses significant clinical utility and offers a valuable reference for future work.

In clinical practice, prognosis and diagnosis are closely linked and often exhibit positive correlations; an experienced clinician typically considers multiple aspects concurrently  \citep{croft2015science}. Modeling a single task tends to overlook the interdependencies among various diagnostic and prognostic factors, making it imperative to incorporate multiple objectives into the modeling process. We have explored classical multi-task learning approaches in experiments. However, performance tends to decline when tasks are learned simultaneously, underscoring the need for more effective multi-task learning algorithms and a more powerful foundation model.

\section{Conclusion}
\label{sec:conclusion}

This paper introduces Multi-OSCC, a novel clinical scenario-oriented multi-task dataset for OSCC diagnosis and prognosis, accompanied by comprehensive benchmarks under single-task and multi-task settings. Our key findings include: (1) introducing pathology-specific pre-training substantially improves both OSCC diagnosis and prognosis performance; 
(2) tumor recurrence prediction is highly sensitive to color variations, with stain normalization improving diagnostic tasks but impairing recurrence prediction, highlighting the need for task-specific preprocessing; 
and (3) while single-task models achieve promising AUC scores, balancing performance across diverse clinical tasks remains challenging for multi-task frameworks, highlighting avenues for future innovation. To encourage further research, we have made the dataset publicly available, paving the way toward improved automated systems for automated clinical evaluation of OSCC.




\bibliographystyle{cas-model2-names}

\bibliography{cas-refs}

\begin{thebibliography}{46}
\expandafter\ifx\csname natexlab\endcsname\relax\def\natexlab#1{#1}\fi
\providecommand{\url}[1]{\texttt{#1}}
\providecommand{\href}[2]{#2}
\providecommand{\path}[1]{#1}
\providecommand{\DOIprefix}{doi:}
\providecommand{\ArXivprefix}{arXiv:}
\providecommand{\URLprefix}{URL: }
\providecommand{\Pubmedprefix}{pmid:}
\providecommand{\doi}[1]{\href{http://dx.doi.org/#1}{\path{#1}}}
\providecommand{\Pubmed}[1]{\href{pmid:#1}{\path{#1}}}
\providecommand{\bibinfo}[2]{#2}
\ifx\xfnm\relax \def\xfnm[#1]{\unskip,\space#1}\fi
\bibitem[{Afify et~al.(2023)Afify, Mohammed and Hassanien}]{afify2023novel}
\bibinfo{author}{Afify, H.M.}, \bibinfo{author}{Mohammed, K.K.}, \bibinfo{author}{Hassanien, A.E.}, \bibinfo{year}{2023}.
\newblock \bibinfo{title}{Novel prediction model on oscc histopathological images via deep transfer learning combined with grad-cam interpretation}.
\newblock \bibinfo{journal}{Biomedical Signal Processing and Control} \bibinfo{volume}{83}, \bibinfo{pages}{104704}.
\bibitem[{Albalawi et~al.(2024)Albalawi, Thakur, Ramakrishna, Bhatia~Khan, SankaraNarayanan, Almarri and Hadi}]{albalawi2024oral}
\bibinfo{author}{Albalawi, E.}, \bibinfo{author}{Thakur, A.}, \bibinfo{author}{Ramakrishna, M.T.}, \bibinfo{author}{Bhatia~Khan, S.}, \bibinfo{author}{SankaraNarayanan, S.}, \bibinfo{author}{Almarri, B.}, \bibinfo{author}{Hadi, T.H.}, \bibinfo{year}{2024}.
\newblock \bibinfo{title}{Oral squamous cell carcinoma detection using efficientnet on histopathological images}.
\newblock \bibinfo{journal}{Frontiers in Medicine} \bibinfo{volume}{10}, \bibinfo{pages}{1349336}.
\bibitem[{Ba et~al.(2016)Ba, Kiros and Hinton}]{ba2016layer}
\bibinfo{author}{Ba, J.L.}, \bibinfo{author}{Kiros, J.R.}, \bibinfo{author}{Hinton, G.E.}, \bibinfo{year}{2016}.
\newblock \bibinfo{title}{Layer normalization}.
\newblock \bibinfo{journal}{arXiv preprint arXiv:1607.06450} .
\bibitem[{Bray et~al.(2024)Bray, Laversanne, Sung, Ferlay, Siegel, Soerjomataram and Jemal}]{bray2024global}
\bibinfo{author}{Bray, F.}, \bibinfo{author}{Laversanne, M.}, \bibinfo{author}{Sung, H.}, \bibinfo{author}{Ferlay, J.}, \bibinfo{author}{Siegel, R.L.}, \bibinfo{author}{Soerjomataram, I.}, \bibinfo{author}{Jemal, A.}, \bibinfo{year}{2024}.
\newblock \bibinfo{title}{Global cancer statistics 2022: Globocan estimates of incidence and mortality worldwide for 36 cancers in 185 countries}.
\newblock \bibinfo{journal}{CA: a cancer journal for clinicians} \bibinfo{volume}{74}, \bibinfo{pages}{229--263}.
\bibitem[{Caron et~al.(2021)Caron, Touvron, Misra, J{\'e}gou, Mairal, Bojanowski and Joulin}]{caron2021emerging}
\bibinfo{author}{Caron, M.}, \bibinfo{author}{Touvron, H.}, \bibinfo{author}{Misra, I.}, \bibinfo{author}{J{\'e}gou, H.}, \bibinfo{author}{Mairal, J.}, \bibinfo{author}{Bojanowski, P.}, \bibinfo{author}{Joulin, A.}, \bibinfo{year}{2021}.
\newblock \bibinfo{title}{Emerging properties in self-supervised vision transformers}, in: \bibinfo{booktitle}{Proceedings of the IEEE/CVF international conference on computer vision}, pp. \bibinfo{pages}{9650--9660}.
\bibitem[{Chattopadhay et~al.(2018)Chattopadhay, Sarkar, Howlader and Balasubramanian}]{chattopadhay2018grad}
\bibinfo{author}{Chattopadhay, A.}, \bibinfo{author}{Sarkar, A.}, \bibinfo{author}{Howlader, P.}, \bibinfo{author}{Balasubramanian, V.N.}, \bibinfo{year}{2018}.
\newblock \bibinfo{title}{Grad-cam++: Generalized gradient-based visual explanations for deep convolutional networks}, in: \bibinfo{booktitle}{2018 IEEE winter conference on applications of computer vision (WACV)}, \bibinfo{organization}{IEEE}. pp. \bibinfo{pages}{839--847}.
\bibitem[{Chaudhary et~al.(2024)Chaudhary, Rai, Rao, Faizan, Augustine, Chaurasia, Mishra, Chandra, Chauhan and Ahmad}]{chaudhary2024high}
\bibinfo{author}{Chaudhary, N.}, \bibinfo{author}{Rai, A.}, \bibinfo{author}{Rao, A.M.}, \bibinfo{author}{Faizan, M.I.}, \bibinfo{author}{Augustine, J.}, \bibinfo{author}{Chaurasia, A.}, \bibinfo{author}{Mishra, D.}, \bibinfo{author}{Chandra, A.}, \bibinfo{author}{Chauhan, V.}, \bibinfo{author}{Ahmad, T.}, \bibinfo{year}{2024}.
\newblock \bibinfo{title}{High-resolution ai image dataset for diagnosing oral submucous fibrosis and squamous cell carcinoma}.
\newblock \bibinfo{journal}{Scientific Data} \bibinfo{volume}{11}, \bibinfo{pages}{1050}.
\bibitem[{Chawla et~al.(2002)Chawla, Bowyer, Hall and Kegelmeyer}]{chawla2002smote}
\bibinfo{author}{Chawla, N.V.}, \bibinfo{author}{Bowyer, K.W.}, \bibinfo{author}{Hall, L.O.}, \bibinfo{author}{Kegelmeyer, W.P.}, \bibinfo{year}{2002}.
\newblock \bibinfo{title}{Smote: synthetic minority over-sampling technique}.
\newblock \bibinfo{journal}{Journal of artificial intelligence research} \bibinfo{volume}{16}, \bibinfo{pages}{321--357}.
\bibitem[{Chen et~al.(2022)Chen, Chen, Li, Chen, Trister, Krishnan and Mahmood}]{chen2022scaling}
\bibinfo{author}{Chen, R.J.}, \bibinfo{author}{Chen, C.}, \bibinfo{author}{Li, Y.}, \bibinfo{author}{Chen, T.Y.}, \bibinfo{author}{Trister, A.D.}, \bibinfo{author}{Krishnan, R.G.}, \bibinfo{author}{Mahmood, F.}, \bibinfo{year}{2022}.
\newblock \bibinfo{title}{Scaling vision transformers to gigapixel images via hierarchical self-supervised learning}, in: \bibinfo{booktitle}{Proceedings of the IEEE/CVF conference on computer vision and pattern recognition}, pp. \bibinfo{pages}{16144--16155}.
\bibitem[{Chen et~al.(2018)Chen, Badrinarayanan, Lee and Rabinovich}]{chen2018gradnorm}
\bibinfo{author}{Chen, Z.}, \bibinfo{author}{Badrinarayanan, V.}, \bibinfo{author}{Lee, C.Y.}, \bibinfo{author}{Rabinovich, A.}, \bibinfo{year}{2018}.
\newblock \bibinfo{title}{Gradnorm: Gradient normalization for adaptive loss balancing in deep multitask networks}, in: \bibinfo{booktitle}{International conference on machine learning}, \bibinfo{organization}{PMLR}. pp. \bibinfo{pages}{794--803}.
\bibitem[{Claridge-Chang and Assam(2016)}]{claridge2016estimation}
\bibinfo{author}{Claridge-Chang, A.}, \bibinfo{author}{Assam, P.N.}, \bibinfo{year}{2016}.
\newblock \bibinfo{title}{Estimation statistics should replace significance testing}.
\newblock \bibinfo{journal}{Nature methods} \bibinfo{volume}{13}, \bibinfo{pages}{108--109}.
\bibitem[{Corredor et~al.(2019)Corredor, Wang, Zhou, Lu, Fu, Syrigos, Rimm, Yang, Romero, Schalper et~al.}]{corredor2019spatial}
\bibinfo{author}{Corredor, G.}, \bibinfo{author}{Wang, X.}, \bibinfo{author}{Zhou, Y.}, \bibinfo{author}{Lu, C.}, \bibinfo{author}{Fu, P.}, \bibinfo{author}{Syrigos, K.}, \bibinfo{author}{Rimm, D.L.}, \bibinfo{author}{Yang, M.}, \bibinfo{author}{Romero, E.}, \bibinfo{author}{Schalper, K.A.}, et~al., \bibinfo{year}{2019}.
\newblock \bibinfo{title}{Spatial architecture and arrangement of tumor-infiltrating lymphocytes for predicting likelihood of recurrence in early-stage non--small cell lung cancer}.
\newblock \bibinfo{journal}{Clinical cancer research} \bibinfo{volume}{25}, \bibinfo{pages}{1526--1534}.
\bibitem[{Crawshaw(2020)}]{crawshaw2020multi}
\bibinfo{author}{Crawshaw, M.}, \bibinfo{year}{2020}.
\newblock \bibinfo{title}{Multi-task learning with deep neural networks: A survey}.
\newblock \bibinfo{journal}{arXiv preprint arXiv:2009.09796} .
\bibitem[{Croft et~al.(2015)Croft, Altman, Deeks, Dunn, Hay, Hemingway, LeResche, Peat, Perel, Petersen et~al.}]{croft2015science}
\bibinfo{author}{Croft, P.}, \bibinfo{author}{Altman, D.G.}, \bibinfo{author}{Deeks, J.J.}, \bibinfo{author}{Dunn, K.M.}, \bibinfo{author}{Hay, A.D.}, \bibinfo{author}{Hemingway, H.}, \bibinfo{author}{LeResche, L.}, \bibinfo{author}{Peat, G.}, \bibinfo{author}{Perel, P.}, \bibinfo{author}{Petersen, S.E.}, et~al., \bibinfo{year}{2015}.
\newblock \bibinfo{title}{The science of clinical practice: disease diagnosis or patient prognosis? evidence about “what is likely to happen” should shape clinical practice}.
\newblock \bibinfo{journal}{BMC medicine} \bibinfo{volume}{13}, \bibinfo{pages}{1--8}.
\bibitem[{Deng et~al.(2009)Deng, Dong, Socher, Li, Li and Fei-Fei}]{deng2009imagenet}
\bibinfo{author}{Deng, J.}, \bibinfo{author}{Dong, W.}, \bibinfo{author}{Socher, R.}, \bibinfo{author}{Li, L.J.}, \bibinfo{author}{Li, K.}, \bibinfo{author}{Fei-Fei, L.}, \bibinfo{year}{2009}.
\newblock \bibinfo{title}{Imagenet: A large-scale hierarchical image database}, in: \bibinfo{booktitle}{2009 IEEE conference on computer vision and pattern recognition}, \bibinfo{organization}{Ieee}. pp. \bibinfo{pages}{248--255}.
\bibitem[{DiCiccio and Efron(1996)}]{diciccio1996bootstrap}
\bibinfo{author}{DiCiccio, T.J.}, \bibinfo{author}{Efron, B.}, \bibinfo{year}{1996}.
\newblock \bibinfo{title}{Bootstrap confidence intervals}.
\newblock \bibinfo{journal}{Statistical science} \bibinfo{volume}{11}, \bibinfo{pages}{189--228}.
\bibitem[{Dosovitskiy et~al.(2020)Dosovitskiy, Beyer, Kolesnikov, Weissenborn, Zhai, Unterthiner, Dehghani, Minderer, Heigold, Gelly et~al.}]{dosovitskiy2020image}
\bibinfo{author}{Dosovitskiy, A.}, \bibinfo{author}{Beyer, L.}, \bibinfo{author}{Kolesnikov, A.}, \bibinfo{author}{Weissenborn, D.}, \bibinfo{author}{Zhai, X.}, \bibinfo{author}{Unterthiner, T.}, \bibinfo{author}{Dehghani, M.}, \bibinfo{author}{Minderer, M.}, \bibinfo{author}{Heigold, G.}, \bibinfo{author}{Gelly, S.}, et~al., \bibinfo{year}{2020}.
\newblock \bibinfo{title}{An image is worth 16x16 words: Transformers for image recognition at scale}.
\newblock \bibinfo{journal}{arXiv preprint arXiv:2010.11929} .
\bibitem[{Fu et~al.(2020)Fu, Chen, Li, Jing, Hu, Liu, Bao, Hong, Shi, Li et~al.}]{fu2020deep}
\bibinfo{author}{Fu, Q.}, \bibinfo{author}{Chen, Y.}, \bibinfo{author}{Li, Z.}, \bibinfo{author}{Jing, Q.}, \bibinfo{author}{Hu, C.}, \bibinfo{author}{Liu, H.}, \bibinfo{author}{Bao, J.}, \bibinfo{author}{Hong, Y.}, \bibinfo{author}{Shi, T.}, \bibinfo{author}{Li, K.}, et~al., \bibinfo{year}{2020}.
\newblock \bibinfo{title}{A deep learning algorithm for detection of oral cavity squamous cell carcinoma from photographic images: A retrospective study}.
\newblock \bibinfo{journal}{EClinicalMedicine} \bibinfo{volume}{27}.
\bibitem[{Glorot et~al.(2011)Glorot, Bordes and Bengio}]{glorot2011deep}
\bibinfo{author}{Glorot, X.}, \bibinfo{author}{Bordes, A.}, \bibinfo{author}{Bengio, Y.}, \bibinfo{year}{2011}.
\newblock \bibinfo{title}{Deep sparse rectifier neural networks}, in: \bibinfo{booktitle}{Proceedings of the fourteenth international conference on artificial intelligence and statistics}, \bibinfo{organization}{JMLR Workshop and Conference Proceedings}. pp. \bibinfo{pages}{315--323}.
\bibitem[{He et~al.(2016)He, Zhang, Ren and Sun}]{he2016deep}
\bibinfo{author}{He, K.}, \bibinfo{author}{Zhang, X.}, \bibinfo{author}{Ren, S.}, \bibinfo{author}{Sun, J.}, \bibinfo{year}{2016}.
\newblock \bibinfo{title}{Deep residual learning for image recognition}, in: \bibinfo{booktitle}{Proceedings of the IEEE conference on computer vision and pattern recognition}, pp. \bibinfo{pages}{770--778}.
\bibitem[{Huang et~al.(2017)Huang, Liu, Van Der~Maaten and Weinberger}]{huang2017densely}
\bibinfo{author}{Huang, G.}, \bibinfo{author}{Liu, Z.}, \bibinfo{author}{Van Der~Maaten, L.}, \bibinfo{author}{Weinberger, K.Q.}, \bibinfo{year}{2017}.
\newblock \bibinfo{title}{Densely connected convolutional networks}, in: \bibinfo{booktitle}{Proceedings of the IEEE conference on computer vision and pattern recognition}, pp. \bibinfo{pages}{4700--4708}.
\bibitem[{Kang et~al.(2023)Kang, Song, Park, Yoo and Pereira}]{kang2023benchmarking}
\bibinfo{author}{Kang, M.}, \bibinfo{author}{Song, H.}, \bibinfo{author}{Park, S.}, \bibinfo{author}{Yoo, D.}, \bibinfo{author}{Pereira, S.}, \bibinfo{year}{2023}.
\newblock \bibinfo{title}{Benchmarking self-supervised learning on diverse pathology datasets}, in: \bibinfo{booktitle}{Proceedings of the IEEE/CVF Conference on Computer Vision and Pattern Recognition}, pp. \bibinfo{pages}{3344--3354}.
\bibitem[{Kayser et~al.(2009)Kayser, Schultz, Goldmann, G{\"o}rtler, Kayser and Vollmer}]{kayser2009theory}
\bibinfo{author}{Kayser, K.}, \bibinfo{author}{Schultz, H.}, \bibinfo{author}{Goldmann, T.}, \bibinfo{author}{G{\"o}rtler, J.}, \bibinfo{author}{Kayser, G.}, \bibinfo{author}{Vollmer, E.}, \bibinfo{year}{2009}.
\newblock \bibinfo{title}{Theory of sampling and its application in tissue based diagnosis}.
\newblock \bibinfo{journal}{Diagnostic Pathology} \bibinfo{volume}{4}, \bibinfo{pages}{1--13}.
\bibitem[{Liu et~al.(2022)Liu, Hu, Lin, Yao, Xie, Wei, Ning, Cao, Zhang, Dong et~al.}]{liu2022swin}
\bibinfo{author}{Liu, Z.}, \bibinfo{author}{Hu, H.}, \bibinfo{author}{Lin, Y.}, \bibinfo{author}{Yao, Z.}, \bibinfo{author}{Xie, Z.}, \bibinfo{author}{Wei, Y.}, \bibinfo{author}{Ning, J.}, \bibinfo{author}{Cao, Y.}, \bibinfo{author}{Zhang, Z.}, \bibinfo{author}{Dong, L.}, et~al., \bibinfo{year}{2022}.
\newblock \bibinfo{title}{Swin transformer v2: Scaling up capacity and resolution}, in: \bibinfo{booktitle}{Proceedings of the IEEE/CVF conference on computer vision and pattern recognition}, pp. \bibinfo{pages}{12009--12019}.
\bibitem[{Liu et~al.(2018)Liu, Shen, Lakshminarasimhan, Liang, Zadeh and Morency}]{liu2018efficient}
\bibinfo{author}{Liu, Z.}, \bibinfo{author}{Shen, Y.}, \bibinfo{author}{Lakshminarasimhan, V.B.}, \bibinfo{author}{Liang, P.P.}, \bibinfo{author}{Zadeh, A.}, \bibinfo{author}{Morency, L.P.}, \bibinfo{year}{2018}.
\newblock \bibinfo{title}{Efficient low-rank multimodal fusion with modality-specific factors}.
\newblock \bibinfo{journal}{arXiv preprint arXiv:1806.00064} .
\bibitem[{Lu et~al.(2024)Lu, Chen, Williamson, Chen, Liang, Ding, Jaume, Odintsov, Le, Gerber et~al.}]{lu2024visual}
\bibinfo{author}{Lu, M.Y.}, \bibinfo{author}{Chen, B.}, \bibinfo{author}{Williamson, D.F.}, \bibinfo{author}{Chen, R.J.}, \bibinfo{author}{Liang, I.}, \bibinfo{author}{Ding, T.}, \bibinfo{author}{Jaume, G.}, \bibinfo{author}{Odintsov, I.}, \bibinfo{author}{Le, L.P.}, \bibinfo{author}{Gerber, G.}, et~al., \bibinfo{year}{2024}.
\newblock \bibinfo{title}{A visual-language foundation model for computational pathology}.
\newblock \bibinfo{journal}{Nature Medicine} \bibinfo{volume}{30}, \bibinfo{pages}{863--874}.
\bibitem[{Lu et~al.(2021)Lu, Williamson, Chen, Chen, Barbieri and Mahmood}]{lu2021data}
\bibinfo{author}{Lu, M.Y.}, \bibinfo{author}{Williamson, D.F.}, \bibinfo{author}{Chen, T.Y.}, \bibinfo{author}{Chen, R.J.}, \bibinfo{author}{Barbieri, M.}, \bibinfo{author}{Mahmood, F.}, \bibinfo{year}{2021}.
\newblock \bibinfo{title}{Data-efficient and weakly supervised computational pathology on whole-slide images}.
\newblock \bibinfo{journal}{Nature biomedical engineering} \bibinfo{volume}{5}, \bibinfo{pages}{555--570}.
\bibitem[{Macenko et~al.(2009)Macenko, Niethammer, Marron, Borland, Woosley, Guan, Schmitt and Thomas}]{macenko2009method}
\bibinfo{author}{Macenko, M.}, \bibinfo{author}{Niethammer, M.}, \bibinfo{author}{Marron, J.S.}, \bibinfo{author}{Borland, D.}, \bibinfo{author}{Woosley, J.T.}, \bibinfo{author}{Guan, X.}, \bibinfo{author}{Schmitt, C.}, \bibinfo{author}{Thomas, N.E.}, \bibinfo{year}{2009}.
\newblock \bibinfo{title}{A method for normalizing histology slides for quantitative analysis}, in: \bibinfo{booktitle}{2009 IEEE international symposium on biomedical imaging: from nano to macro}, \bibinfo{organization}{IEEE}. pp. \bibinfo{pages}{1107--1110}.
\bibitem[{McKinney et~al.(2020)McKinney, Sieniek, Godbole, Godwin, Antropova, Ashrafian, Back, Chesus, Corrado, Darzi et~al.}]{mckinney2020international}
\bibinfo{author}{McKinney, S.M.}, \bibinfo{author}{Sieniek, M.}, \bibinfo{author}{Godbole, V.}, \bibinfo{author}{Godwin, J.}, \bibinfo{author}{Antropova, N.}, \bibinfo{author}{Ashrafian, H.}, \bibinfo{author}{Back, T.}, \bibinfo{author}{Chesus, M.}, \bibinfo{author}{Corrado, G.S.}, \bibinfo{author}{Darzi, A.}, et~al., \bibinfo{year}{2020}.
\newblock \bibinfo{title}{International evaluation of an ai system for breast cancer screening}.
\newblock \bibinfo{journal}{Nature} \bibinfo{volume}{577}, \bibinfo{pages}{89--94}.
\bibitem[{Nechaev et~al.(2024)Nechaev, Pchelnikov and Ivanova}]{nechaev2024hibou}
\bibinfo{author}{Nechaev, D.}, \bibinfo{author}{Pchelnikov, A.}, \bibinfo{author}{Ivanova, E.}, \bibinfo{year}{2024}.
\newblock \bibinfo{title}{Hibou: A family of foundational vision transformers for pathology}.
\newblock \bibinfo{journal}{arXiv preprint arXiv:2406.05074} .
\bibitem[{Oquab et~al.(2023)Oquab, Darcet, Moutakanni, Vo, Szafraniec, Khalidov, Fernandez, Haziza, Massa, El-Nouby et~al.}]{oquab2023dinov2}
\bibinfo{author}{Oquab, M.}, \bibinfo{author}{Darcet, T.}, \bibinfo{author}{Moutakanni, T.}, \bibinfo{author}{Vo, H.}, \bibinfo{author}{Szafraniec, M.}, \bibinfo{author}{Khalidov, V.}, \bibinfo{author}{Fernandez, P.}, \bibinfo{author}{Haziza, D.}, \bibinfo{author}{Massa, F.}, \bibinfo{author}{El-Nouby, A.}, et~al., \bibinfo{year}{2023}.
\newblock \bibinfo{title}{Dinov2: Learning robust visual features without supervision}.
\newblock \bibinfo{journal}{arXiv preprint arXiv:2304.07193} .
\bibitem[{Rahman et~al.(2020)Rahman, Mahanta, Das and Sarma}]{rahman2020histopathological}
\bibinfo{author}{Rahman, T.Y.}, \bibinfo{author}{Mahanta, L.B.}, \bibinfo{author}{Das, A.K.}, \bibinfo{author}{Sarma, J.D.}, \bibinfo{year}{2020}.
\newblock \bibinfo{title}{Histopathological imaging database for oral cancer analysis}.
\newblock \bibinfo{journal}{Data in brief} \bibinfo{volume}{29}, \bibinfo{pages}{105114}.
\bibitem[{Reinhard et~al.(2001)Reinhard, Adhikhmin, Gooch and Shirley}]{reinhard2001color}
\bibinfo{author}{Reinhard, E.}, \bibinfo{author}{Adhikhmin, M.}, \bibinfo{author}{Gooch, B.}, \bibinfo{author}{Shirley, P.}, \bibinfo{year}{2001}.
\newblock \bibinfo{title}{Color transfer between images}.
\newblock \bibinfo{journal}{IEEE Computer graphics and applications} \bibinfo{volume}{21}, \bibinfo{pages}{34--41}.
\bibitem[{Ren et~al.(2020)Ren, Qi, Yuan, Duan and Tao}]{ren2020machine}
\bibinfo{author}{Ren, J.}, \bibinfo{author}{Qi, M.}, \bibinfo{author}{Yuan, Y.}, \bibinfo{author}{Duan, S.}, \bibinfo{author}{Tao, X.}, \bibinfo{year}{2020}.
\newblock \bibinfo{title}{Machine learning--based mri texture analysis to predict the histologic grade of oral squamous cell carcinoma}.
\newblock \bibinfo{journal}{American Journal of Roentgenology} \bibinfo{volume}{215}, \bibinfo{pages}{1184--1190}.
\bibitem[{Sener and Koltun(2018)}]{sener2018multi}
\bibinfo{author}{Sener, O.}, \bibinfo{author}{Koltun, V.}, \bibinfo{year}{2018}.
\newblock \bibinfo{title}{Multi-task learning as multi-objective optimization}.
\newblock \bibinfo{journal}{Advances in neural information processing systems} \bibinfo{volume}{31}.
\bibitem[{Spearman(1961)}]{spearman1961proof}
\bibinfo{author}{Spearman, C.}, \bibinfo{year}{1961}.
\newblock \bibinfo{title}{The proof and measurement of association between two things.} .
\bibitem[{Srivastava et~al.(2014)Srivastava, Hinton, Krizhevsky, Sutskever and Salakhutdinov}]{srivastava2014dropout}
\bibinfo{author}{Srivastava, N.}, \bibinfo{author}{Hinton, G.}, \bibinfo{author}{Krizhevsky, A.}, \bibinfo{author}{Sutskever, I.}, \bibinfo{author}{Salakhutdinov, R.}, \bibinfo{year}{2014}.
\newblock \bibinfo{title}{Dropout: a simple way to prevent neural networks from overfitting}.
\newblock \bibinfo{journal}{The journal of machine learning research} \bibinfo{volume}{15}, \bibinfo{pages}{1929--1958}.
\bibitem[{Stirling et~al.(2021)Stirling, Swain-Bowden, Lucas, Carpenter, Cimini and Goodman}]{stirling2021cellprofiler}
\bibinfo{author}{Stirling, D.R.}, \bibinfo{author}{Swain-Bowden, M.J.}, \bibinfo{author}{Lucas, A.M.}, \bibinfo{author}{Carpenter, A.E.}, \bibinfo{author}{Cimini, B.A.}, \bibinfo{author}{Goodman, A.}, \bibinfo{year}{2021}.
\newblock \bibinfo{title}{Cellprofiler 4: improvements in speed, utility and usability}.
\newblock \bibinfo{journal}{BMC bioinformatics} \bibinfo{volume}{22}, \bibinfo{pages}{1--11}.
\bibitem[{Vahadane et~al.(2016)Vahadane, Peng, Sethi, Albarqouni, Wang, Baust, Steiger, Schlitter, Esposito and Navab}]{vahadane2016structure}
\bibinfo{author}{Vahadane, A.}, \bibinfo{author}{Peng, T.}, \bibinfo{author}{Sethi, A.}, \bibinfo{author}{Albarqouni, S.}, \bibinfo{author}{Wang, L.}, \bibinfo{author}{Baust, M.}, \bibinfo{author}{Steiger, K.}, \bibinfo{author}{Schlitter, A.M.}, \bibinfo{author}{Esposito, I.}, \bibinfo{author}{Navab, N.}, \bibinfo{year}{2016}.
\newblock \bibinfo{title}{Structure-preserving color normalization and sparse stain separation for histological images}.
\newblock \bibinfo{journal}{IEEE transactions on medical imaging} \bibinfo{volume}{35}, \bibinfo{pages}{1962--1971}.
\bibitem[{Vaswani et~al.(2017)Vaswani, Shazeer, Parmar, Uszkoreit, Jones, Gomez, Kaiser and Polosukhin}]{vaswani2017attention}
\bibinfo{author}{Vaswani, A.}, \bibinfo{author}{Shazeer, N.}, \bibinfo{author}{Parmar, N.}, \bibinfo{author}{Uszkoreit, J.}, \bibinfo{author}{Jones, L.}, \bibinfo{author}{Gomez, A.N.}, \bibinfo{author}{Kaiser, {\L}.}, \bibinfo{author}{Polosukhin, I.}, \bibinfo{year}{2017}.
\newblock \bibinfo{title}{Attention is all you need}.
\newblock \bibinfo{journal}{Advances in neural information processing systems} \bibinfo{volume}{30}.
\bibitem[{Vollmer et~al.(2024)Vollmer, Hartmann, Vollmer, Shavlokhova, Brands, K{\"u}bler, Wollborn, Hassel, Couillard-Despres, Lang et~al.}]{vollmer2024multimodal}
\bibinfo{author}{Vollmer, A.}, \bibinfo{author}{Hartmann, S.}, \bibinfo{author}{Vollmer, M.}, \bibinfo{author}{Shavlokhova, V.}, \bibinfo{author}{Brands, R.C.}, \bibinfo{author}{K{\"u}bler, A.}, \bibinfo{author}{Wollborn, J.}, \bibinfo{author}{Hassel, F.}, \bibinfo{author}{Couillard-Despres, S.}, \bibinfo{author}{Lang, G.}, et~al., \bibinfo{year}{2024}.
\newblock \bibinfo{title}{Multimodal artificial intelligence-based pathogenomics improves survival prediction in oral squamous cell carcinoma}.
\newblock \bibinfo{journal}{Scientific reports} \bibinfo{volume}{14}, \bibinfo{pages}{5687}.
\bibitem[{Warin and Suebnukarn(2024)}]{warin2024deep}
\bibinfo{author}{Warin, K.}, \bibinfo{author}{Suebnukarn, S.}, \bibinfo{year}{2024}.
\newblock \bibinfo{title}{Deep learning in oral cancer-a systematic review}.
\newblock \bibinfo{journal}{BMC Oral Health} \bibinfo{volume}{24}, \bibinfo{pages}{212}.
\bibitem[{Yan et~al.(2024)Yan, Sun, Jin, Liu, He, Guan and Chen}]{yan2024shapley}
\bibinfo{author}{Yan, R.}, \bibinfo{author}{Sun, Q.}, \bibinfo{author}{Jin, C.}, \bibinfo{author}{Liu, Y.}, \bibinfo{author}{He, Y.}, \bibinfo{author}{Guan, T.}, \bibinfo{author}{Chen, H.}, \bibinfo{year}{2024}.
\newblock \bibinfo{title}{Shapley values-enabled progressive pseudo bag augmentation for whole-slide image classification}.
\newblock \bibinfo{journal}{IEEE Transactions on Medical Imaging} .
\bibitem[{Zhou et~al.(2021)Zhou, Wei, Wang, Shen, Xie, Yuille and Kong}]{zhou2021ibot}
\bibinfo{author}{Zhou, J.}, \bibinfo{author}{Wei, C.}, \bibinfo{author}{Wang, H.}, \bibinfo{author}{Shen, W.}, \bibinfo{author}{Xie, C.}, \bibinfo{author}{Yuille, A.}, \bibinfo{author}{Kong, T.}, \bibinfo{year}{2021}.
\newblock \bibinfo{title}{ibot: Image bert pre-training with online tokenizer}.
\newblock \bibinfo{journal}{arXiv preprint arXiv:2111.07832} .
\bibitem[{Zhou et~al.(2024)Zhou, Wu, Hong, Huang, Jia, Lu, Cheng, Xu, Yang and Wu}]{zhou2024pathology}
\bibinfo{author}{Zhou, J.}, \bibinfo{author}{Wu, H.}, \bibinfo{author}{Hong, X.}, \bibinfo{author}{Huang, Y.}, \bibinfo{author}{Jia, B.}, \bibinfo{author}{Lu, J.}, \bibinfo{author}{Cheng, B.}, \bibinfo{author}{Xu, M.}, \bibinfo{author}{Yang, M.}, \bibinfo{author}{Wu, T.}, \bibinfo{year}{2024}.
\newblock \bibinfo{title}{A pathology-based diagnosis and prognosis intelligent system for oral squamous cell carcinoma using semi-supervised learning}.
\newblock \bibinfo{journal}{Expert Systems with Applications} \bibinfo{volume}{254}, \bibinfo{pages}{124242}.
\bibitem[{Zuley et~al.(2016)Zuley, Jarosz, Kirk, Lee, Colen, Garcia, Delbeke, Pham, Nagy, Sevinc et~al.}]{zuley2016cancer}
\bibinfo{author}{Zuley, M.}, \bibinfo{author}{Jarosz, R.}, \bibinfo{author}{Kirk, S.}, \bibinfo{author}{Lee, Y.}, \bibinfo{author}{Colen, R.}, \bibinfo{author}{Garcia, K.}, \bibinfo{author}{Delbeke, D.}, \bibinfo{author}{Pham, M.}, \bibinfo{author}{Nagy, P.}, \bibinfo{author}{Sevinc, G.}, et~al., \bibinfo{year}{2016}.
\newblock \bibinfo{title}{The cancer genome atlas head-neck squamous cell carcinoma collection (tcga-hnsc)}.
\newblock \bibinfo{journal}{The Cancer Imaging Archive} .

\end{thebibliography}



\end{document}